\documentstyle[preprint,aps,epsfig]{revtex}
\begin{document}

\tighten
\draft
\preprint{
\vbox{
\hbox{March 1999}
\hbox{ADP-99-14/T357}
}}

\title{Charm in the Nucleon}
\author{F.M. Steffens}
\address{Instituto de Fisica -- USP,
	C.P. 66 318, 05315-970,
	SP, Brazil}
\author{W. Melnitchouk}
\address{Institut f\"ur Kernphysik,
	Forschungszentrum J\"ulich,
	D-52425 J\"ulich, Germany}
\author{A.W. Thomas}
\address{Department of Physics and Mathematical Physics,
	and Special Research Center for the Subatomic
	Structure of Matter,
	University of Adelaide, 5005,
	Australia}
\maketitle

\begin{abstract}
A next-to-leading order analysis of inelastic electroproduction of
charm is performed using an interpolating scheme which maps smoothly
onto massless QCD evolution at large $Q^2$ and photon--gluon fusion
at small $Q^2$.
In contrast with earlier analyses, this scheme allows the inclusion
of quark and target mass effects and heavy quark thresholds, as well
as possible non-perturbative, or intrinsic, charm contributions.
We find no conclusive evidence in favor of an intrinsic charm
component in the nucleon, although several data points which
disagree with perturbative QCD expectations will need to be
checked by future experiments.
\end{abstract}

\section{Introduction}

Understanding the role played in the nucleon by heavy quarks, such as
the charm quark, is necessary for a number of reasons.
Firstly, one cannot claim to have unraveled the rich structure of the
nucleon sea until one has mapped out the details of the distribution
of its virtual charm and heavier\footnote{Although in practice the direct
accessibility of bottom and top quark densities is likely to remain
elusive for some time.} flavors.
Secondly, in the absence of a direct probe of gluons, charm
leptoproduction remains one of the main sources of information
on the nucleon's {\em gluon} distribution.
Furthermore, tagging charm in neutrino and antineutrino scattering
allows one to probe the strange and antistrange quark densities in
the nucleon.
{}From a more theoretical point of view, in order to have a reliable
procedure through which to analyze deep-inelastic scattering data,
one needs to consistently incorporate heavy quark masses and threshold
effects in the QCD evolution equations.

Recently there have been important new data on the charm structure
function, $F_2^c$, of the proton from the H1 \cite{H1} and ZEUS
\cite{ZEUS} Collaborations at HERA, which have probed the small-$x$
region down to $x = 8 \times 10^{-4}$ and $2 \times 10^{-4}$,
respectively.
At these values of $x$, the charm contribution to the total proton
structure function, $F_2^p$, is found to be around 25\%, which is a
considerably larger fraction than that found by the European Muon
Collaboration at CERN \cite{EMC} at somewhat larger $x$, where it
was only $\sim 1\%$ of $F_2^p$.
Extensive theoretical analyses in recent years have generally
served to confirm that the bulk of the $F_2^c$ data can be
described through perturbative generation of charm within QCD.

At the same time, there are several lingering pieces of data
which seem to suggest the possibility that a small component
of charm exists which is intrinsically non-perturbative in origin
\cite{BROD,HM,HSV,IT,GOL}.
One of these is the EMC data \cite{EMC} at large $x$,
some of which appear to lie above the perturbative QCD predictions,
and which have in the past been taken \cite{BROD} as evidence in
favor of a non-perturbative, or ``intrinsic'' charm component.
Furthermore, as recently discussed in Ref.\cite{HSV}, there are
some indications of intrinsic charm also from hadronic reactions,
such as leading charm production in $\pi N$ and $Y N$ scattering.
As found in Ref.\cite{VB}, some intrinsic charm may account for
the larger than expected number of fast correlated $J/\psi$ pairs
seen in the NA3 $\pi N$ experiment at CERN \cite{NA3}, as well as
for the anomalous polarization of $J/\psi$ seen in inclusive
$J/\psi$ production in $\pi N$ collisions \cite{POL}.

The initial analyses of the leptoproduction data suggested an
intrinsic charm component of the nucleon with normalization of
around 1\%, although subsequent, more sophisticated, treatments
incorporating higher-order effects tended to find only a fraction
of a percent.
The basic starting assumption in these analyses is that the
perturbative generation of charm proceeds through the photon--gluon
fusion process, illustrated in Figure~1.
In their pioneering analysis, Hoffman and Moore \cite{HM}
considered the ${\cal O}(\alpha_s)$ corrections to the intrinsic
charm distributions, as well as quark and target mass effects on
the charm cross section.
With these refinements, their work suggested that an intrinsic charm
component of the proton, at a level of order 0.3\%, was allowed
--- even required --- by the data.
In a more recent study, Harris, Smith and Vogt \cite{HSV}
reanalyzed the EMC data within the framework of Ref.~\cite{HM}
with updated parton distributions, and ${\cal O}(\alpha_s$)
corrections to the hard scattering cross section.
They found that while no clear statements about intrinsic charm
could be made at lower energy transfers, $\nu \alt 100$ GeV,
data at larger $\nu \sim 170$ GeV could best be fitted with an
additional $0.86 \pm 0.60\%$ intrinsic charm component \cite{HSV}.

Despite these claims of evidence for intrinsic charm, a number
of recent global parameterizations of parton distributions
\cite{MRST,GRV} have concluded that all $F_2^c$ data, including
those from the EMC, can be understood in terms of perturbative
QCD alone, with no additional intrinsic charm component necessary.
In view of these conflicting results, it seems important 
that the question of the presence or otherwise of intrinsic charm
be addressed using the latest available theoretical techniques
for calculating $F_2^c$, as well as different models for intrinsic
charm.

To set the context for our analysis, we should point out a number
of features which have been common to all earlier treatments of
intrinsic charm in leptoproduction.
Firstly, one invariably assumes that photon--gluon fusion
(together with perturbative QCD corrections) is solely responsible
for perturbatively generated charm, {\it irrespective} of the
scale $Q^2$ at which one works.
In light of recent developments in incorporating quark mass
effects in perturbative QCD, it is now in principle possible
to go beyond this approximation and include more appropriate
treatments in kinematic regions where photon--gluon fusion
alone may not be the only relevant perturbative mechanism.
A second point concerns the way that intrinsic charm is combined
with the perturbative contributions.
In particular, no treatment to date has considered the effects
of $Q^2$ evolution of the non-perturbative charm distribution
from the ``intrinsic'' scale to the scale where data are taken.

In this paper we perform a comprehensive next-to-leading order   
analysis of the charm production data in which we compare several different
methods of implementing charm in QCD structure functions.
In particular, we consider:

$\bullet$ {\em Massless Evolution.}
This approach assumes that below a certain scale, $\mu^2$,
there are $n_f$ active flavors.
Above this scale, one switches the number of flavors to
$n_f + 1$ in the coupling constant, coefficient functions
and in the parton distributions.
This scheme is sometimes referred to as the
Variable Flavor Number Scheme (VFNS).

$\bullet$ {\em Photon--Gluon Fusion.}
In this scheme, the heavy quark is never treated as a parton,
in the sense that the number of active flavors does not change,
and a quark distribution for the heavy quark is never introduced.
The entire contribution to the physical structure function from
heavy quarks is generated through the photon--gluon fusion process
depicted in Figure~1.
This scheme is usually referred to as the
Fixed Flavor Number Scheme (FFNS).

$\bullet$ {\em Interpolating Schemes.}
While photon--gluon fusion should provide a reliable description
of charm production in the region close to $m_c^2$, it certainly
should fail at large scales because of the lack of resummation
of the logarithms in $m_c^2$.
Conversely, massless evolution should be good at large $Q^2$,
but not in the region around $m_c^2$.
Schemes such as that proposed by Aivazis et al. \cite{ACOT} reproduce
the relevant features of both the VFNS and FFNS approaches ---
namely, a reduction to FFNS when $Q^2 \sim m_c^2$, and to VFNS
when $Q^2 \gg m_c^2$.
Unlike Ref.\cite{ACOT}, however, we include the full quark mass
dependence not only in the gluon sector, but also in the quark
sector, which has become possible since the recent work of Kretzer
and Schienbein \cite{KRETZER}.

Throughout the analysis we shall work in the $\overline{\rm MS}$
scheme, and introduce an explicit charm distribution at a scale
$m_c^2$ for the VFNS and interpolating schemes.
Furthermore, our philosophy is that if there is any intrinsic
charm at all in the nucleon, it should be considered as a parton
distribution for all scales $Q^2 \ge m_c^2$, if one is to use
$\overline{\rm MS}$ anomalous dimensions when evolving the
distributions.

In the following sections we describe in turn the three schemes
for incorporating charm, beginning with the standard massless
QCD evolution and the photon--gluon fusion process in Sections
II and III, respectively.
The interpolating scheme connecting these two limiting cases
is described in Section IV.
Incorporation of intrinsic charm into the analysis is outlined in
Section V, where we survey several models for the non-perturbative
charm component.
Finally, in Section VI some conclusions are drawn from the analysis.

\newpage
\section{Massless Evolution (VFNS)}

The simplest approach to describing charm in the nucleon is to assume
that the charm distribution is $c(x,\mu^2) = \bar c(x,\mu^2) = 0$
below a certain scale, $\mu$, and evolve charm quarks as massless
partons.
Indeed, until recently, this has been the standard approach adopted
in global analyses of parton distribution functions.
In a scheme such as $\overline{\rm MS}$, the scale $\mu$ is taken to
be the charm quark mass, $m_c$ \cite{TUNG}.
Above threshold, $Q^2 > m_c^2$, the number of active flavors increases
by one in the coupling constant, splitting and coefficient functions.
In this way the charm quark distribution is generated solely through
radiative corrections from gluons and light quarks\footnote{
Just as for the $c$ quark, the $u$, $d$, and $s$ quarks are light,
so too for heavier quarks, such as $b$, the $u$, $d$, $s$ and $c$
quarks would be light, and so on.} present at $Q^2 < m_c^2$.

At next-to-leading order, the charm quark contribution to the
proton structure function is given by:
\begin{eqnarray}
\label{massless}
F_2^{c\ ({\rm VFNS})}(x,Q^2)
&=& e_c^2\ C_q(x,Q^2) \otimes \left( x c(x,Q^2) + x \bar c(x,Q^2) \right)
 +  e_c^2\ C_g(x,Q^2) \otimes x g(x,Q^2)
\end{eqnarray}
with the boundary condition $c(x,m_c^2)=\bar c(x,m_c^2)=0$.
The convolution ``$\otimes$'' here is defined by
$f(x) \otimes h(x) \equiv \int_x^1 dy f(x/y) h(y)$.
The quark and gluon coefficient functions are those for
massless quarks:
\begin{mathletters}
\label{e1}
\begin{eqnarray}
\label{e1a}
C_q(x,Q^2)
&=& \delta(1-x)
 +  \frac{\alpha_s(Q^2)}{4 \pi}
    C_F \left\{ 4 \left(\frac{\ln(1 - x)}{1 - x}\right)_+
	       - \frac{3}{(1 - x)_+}
	       - 2 (1 + x) \ln(1 - x)
	\right.							\nonumber\\
& & - \left. 2 \frac{1 + x^2}{1 - x} \ln x  +  6  +  4 x
      \right\},							\\
\label{e1b}
C_g(x,Q^2)
&=& \frac{\alpha_s(Q^2)}{4 \pi}
    4 T_F \left\{ (1 - 2 x + 2 x^2)\ln\frac{1 - x}{x} - 4 + 32 x (1 - x)
	  \right\},
\end{eqnarray}
\end{mathletters}%
where the $(1-x)_+$ refers to the standard ``+''prescription
\cite{PLUS}.

The gluon coefficient function, $C_g$, is a pure singlet quantity,
and as such is proportional to the number of active flavors.
Since one is calculating the contribution to $F_2^p$ from a single
flavor, $C_g$ defined in Eq.(\ref{e1b}) is the usual gluon
coefficient function divided by the number of active flavors.
The gluon distribution is evolved by taking as input a gluon
distribution function at a scale $m_c^2$, together with a quark
singlet distribution,
$\Sigma(x,m_c^2)
= \sum_{q=u,d,s,c} \left( q(x,m_c^2) + \bar q(x,m_c^2) \right)$.
Using the boundary condition $c(x,m_c^2)=\bar c(x,m_c^2)=0$,
the evolution is performed with 4 flavors.

The evolution of the charm quark density is slightly more involved.
Firstly, one evolves the singlet combination from $m_c^2$ to $Q^2$
using the full number of active flavors.
The quark non-singlet combination,
$\eta(x,Q^2)
= \sum_{q=u,d,s} \left( q(x,Q^2) + \bar q(x,Q^2) \right)
- 3 \left( c(x,Q^2) + \bar c(x,Q^2) \right)$,
is then evolved over the same range.
Since $c(x,m_c^2)=\bar c(x,m_c^2)=0$, it turns out that one
is in practice evolving
$\sum_{q=u,d,s} \left( q(x,m_c^2) + \bar q(x,m_c^2) \right)$
first as a singlet and then as a non-singlet.
The charm distribution at $Q^2$ is then given by:
\begin{eqnarray}
\label{e2}
c(x,Q^2) + \bar c(x,Q^2)
&=& {1 \over 4} \left( \Sigma(x,Q^2) - \eta(x,Q^2) \right).
\end{eqnarray}

For large $Q^2$, $Q^2 \gg m_c^2$, one expects the generation
of charm through massless evolution to be a good approximation,
a result confirmed in previous studies \cite{BUZA}.
However, at smaller $Q^2$, $Q^2 \sim m_c^2$, mass effects become
important, and this approach eventually breaks down.
In this region a better approximation is the photon--gluon fusion
process, which we discuss next.

\section{Photon--Gluon Fusion (FFNS)}

In the photon--gluon fusion approach the charm quark is treated
not as a parton, but rather as a heavy quark, so that a charm
quark distribution is never explicitly introduced.
Hence, $c=\bar c=0$ for all $x$ and $Q^2$.
The charm structure function is given directly by a convolution
of the gluon distribution and the hard photon--gluon cross section,
Figure~1.

At lowest order, the only process producing a heavy $q \bar q$
pair is the photon--gluon fusion, and the structure function is
given by:
\begin{eqnarray}
\label{pgf}
F_2^{c\ ({\rm FFNS})}(x,Q^2)
&=& e_c^2\ \frac{\alpha_s(\mu^2)}{4 \pi}
\int_x^{z_{max}} dz\
H_g(z,Q^2,m_c^2)\ {x \over z} g\left({x\over z},\mu^2\right),
\end{eqnarray}
where $z_{max} = 1/(1 + 4m_c^2/Q^2)$ and $\mu^2$ is the mass
factorization (and renormalization) scale.
The partonic cross section for producing a massive quark pair
is given by:
\begin{eqnarray}
H_g(z,Q^2,m_c^2) 
&=& 4 \left( 1 - 2 z + 2 z^2 + 4 z (1 - 3 z)\frac{m_c^2}{Q^2} 
	       - 8 z^2 \frac{m_c^4}{Q^4}
      \right)
\ln\left(\frac{1 + \beta}{1 - \beta}\right)		\nonumber\\
&+& \beta \left( 32 z (1 - z) - 4
	       - 16 z (1 - z) \frac{m_c^2}{Q^2}
	  \right),
\end{eqnarray}
with
\begin{eqnarray}
\beta &=& \sqrt{1 - \frac{4 z}{(1 - z)}\frac{m_c^2}{Q^2}}. 
\end{eqnarray}
Note that the coupling constant is calculated at $\mu^2$ with
3 light flavors, the same number as used throughout in $H_g$.
There is no evolution equation in this approach --- the distribution
$g(x,\mu^2)$ is never evolved.
Evolution equations are introduced only when the logs in $Q^2/m_c^2$
are resummed via the renormalization group equations.
In that case a heavy quark distribution is introduced and one
recovers Eq.(\ref{massless}).

One should mention that the gluon distribution in Eq.(\ref{pgf})
is that given by a leading-order analysis of the data.
This is because in Eq.(\ref{pgf}) one has terms of the form
$\alpha_s(\mu^2) \ln(Q^2/m_c^2)\ g(x,\mu^2)$, which are typical
next-to-leading order contributions if the gluon distribution
is calculated in leading order --- hence one is calculating 
${\cal O}(\alpha_s)$ corrections to the structure function.
The use of a next-to-leading order gluon would produce
next-to-next-to-leading order corrections, which would then
require calculation of radiative corrections to the photon--gluon
fusion term, and the addition of extra terms in Eq.(\ref{pgf})
arising from heavy pair creation from light quarks \cite{HARRIS}.

The advantage of the photon--gluon fusion formulation is its
simplicity.
However, the use of Eq.(\ref{pgf}) could be problematic at very
large $Q^2$ because of the large $\ln(Q^2/m_c^2)$ term appearing
in $H_g$.
One way of circumventing this difficulty is to construct a scheme which
retains the features of photon--gluon fusion at moderate $Q^2$,
but maps onto massless evolution when $Q^2$ becomes very large.

\section{Interpolating Scheme}

The description of charm production from the nucleon at an
arbitrary scale requires a scheme which consistently interpolates
between the two limits of VFNS and FFNS.
A number of authors have pursued the construction of such schemes,
both for unpolarized \cite{ACOT,BUZA,MRRS,THORNE} and 
polarized \cite{STEFFENS} scattering.

With the exception of Ref.\cite{MRRS}, where the splitting functions
were modified for quark mass effects, the essence of the interpolating
schemes is to incorporate a mass dependence, as a function of
$m_c^2/Q^2$, in the coefficient functions which would enable 
Eq.(\ref{massless}) to be recovered when $Q^2 \gg m_c^2$,
and Eq.(\ref{pgf}) when $Q^2 \sim m_c^2$.
By keeping the evolution equations unmodified, one ensures that
the parton distributions (for the 3 light quarks, for the gluon,
and for the newly introduced charm distribution) are defined
consistently in the $\overline{\rm MS}$ scheme.

Until recently, quark mass corrections were only included
in the gluon coefficient function, $H_g$, while the quark
coefficient function was taken to have the form derived for 
massless quarks \cite{MRST,ACOT}.
In fact, it is quite natural to think that the gluon sector,
through the $P_{cg}$ splitting function, would be responsible
for the generation and evolution of a heavy quark distribution.
Furthermore, any quark mass correction (other than that appearing
in the new, massive, scaling variable) would appear in the
${\cal O}(\alpha_s)$ coefficient functions only.
In Ref.\cite{KRETZER}, Kretzer and Schienbein also calculated
the quark mass corrections to the quark coefficient function.
They found that these were small in absolute terms, but of a
similar order of magnitude as the total $F_2^c$ in some
regions of $x$, and for some choices of factorization scale.
There are at least two reasons, therefore, why one needs to
include these effects in the present analysis.
Firstly, with the introduction of intrinsic charm the quark sector
may become relatively more important than the gluon sector in
the region of $x$ where intrinsic charm is non-zero;
and secondly, the analysis of Ref.\cite{KRETZER} focused
on quark mass corrections to the quark coefficient functions
at relatively small values of $x$, while here we also need to
examine the size of these corrections at large $x$.

Of course, the full massive coefficient functions introduce
extra mass logarithms in the quark and gluon sectors,
resulting in double counting when using the evolution
equations for the charm and gluon densities.
This is reminiscent of a scenario where logarithms have not
yet been resummed by the renormalization group equations.
The problem can be circumvented easily enough by requiring
that in the limit $Q^2 \rightarrow \infty$ the massless
coefficient functions are recovered after introducing
subtraction terms $H_{q,g}^{\rm sub}$:
\begin{mathletters}
\label{e3}
\begin{eqnarray}
\label{e3a}
&& \lim_{Q^2 \rightarrow \infty}
\left( H_q(z,Q^2,m_c^2) - H^{\rm sub}_q(z,Q^2,m_c^2)
\right) = C_q (z,Q^2),					\\
\label{e3b}
&& \lim_{Q^2 \rightarrow \infty}
\left( H_g(z,Q^2,m_c^2) - H^{\rm sub}_g(z,Q^2,m_c^2)
\right) = C_g (z,Q^2),
\end{eqnarray}
\end{mathletters}%
where the factorization scale is chosen to be $\mu^2 = Q^2$
for both the interpolating scheme and the VFNS (a choice to 
which we restrict ourselves throughout this analysis).
The function $H_q(z,Q^2,m_c^2)$ in Eq.(\ref{e3a}) is the full
massive quark coefficient function at ${\cal O}(\alpha_s)$,
calculated in Ref.\cite{KRETZER}.
The subtraction terms are given by:
\begin{mathletters}
\label{e4}
\begin{eqnarray}
H_q^{\rm sub}(z,Q^2,m_c^2)
&=& \frac{\alpha_s (Q^2)}{2 \pi}
    C_F \left[ \frac{1 + z^2}{1 - z}
	       \left( \ln\frac{Q^2}{m_c^2} - 1 - 2 \ln(1 - z)
	       \right)
	\right]_+\ ,					\\
H_g^{\rm sub}(z,Q^2,m_c^2)
&=& \frac{\alpha_s (Q^2)}{\pi}
    T_F\ \ln\frac{Q^2}{m_c^2}\ \left( 1 - 2 z + 2 z^2 \right).
\end{eqnarray}
\end{mathletters}
Finally, one has:
\begin{eqnarray}
\label{interp}
F_2^c(x,Q^2)
&=& e_c^2 \int_\xi^1\ dz\
\left( H_q(z,Q^2,m_c^2) - H_q^{\rm sub}(z,Q^2,m_c^2)
\right)
{\xi \over z}
\left( c(\xi/z,Q^2) + \bar c(\xi/z,Q^2) \right)		\nonumber\\
&+& e_c^2 \int_x^{z_{max}}\ dz\
    H_g(z,Q^2,m_c^2)\ {x \over z} g(x/z,Q^2)		\nonumber\\
&-& e_c^2 \int_\xi^1\ dz\
    H_g^{\rm sub}(z,Q^2,m_c^2)\ {\xi \over z} g(\xi/z, Q^2),
\end{eqnarray}
where the scaling variable $\xi$ includes quark mass corrections,
\begin{eqnarray}
\xi &=& {1 \over 2} x \left( 1 + \sqrt{1 + {4 m_c^2\over Q^2}} \right).
\end{eqnarray}
The charm and gluon distributions in Eq.(\ref{interp})
are determined from evolution in exactly the same way
as in the massless case described in Section II.

Figure~\ref{C2} shows $F_2^c$ evaluated according to the
three schemes described above at $x=0.05$ and 0.2, for a 
$Q^2$ range relevant to the EMC data.
In addition, for the interpolating scheme we also show the 
effect of neglecting the quark mass dependence in the quark
coefficient function, so that $H_q$--$H_q^{\rm sub}$ is
replaced by $C_q$ in Eq.(\ref{interp}).
For the parton distributions the GRV parameterizations
\cite{GRV} are used.
Note, however, that in the GRV fit a charm quark distribution
is never introduced explicitly, rather $F_2^c$ is always
calculated via the FFNS.
The charm density is generated from the GRV distributions by
evolving with 3 flavors from $\mu^2=0.4$ GeV$^2$ to $m_c^2$
in next-to-leading order, then from $m_c^2$ to $Q^2$ with 4
flavors according to the VFNS.
%
%
For the FFNS calculation, the gluon distribution is evolved in
leading order from $\mu^2=0.26$ GeV$^2$ to $m_c^2$ with 3 flavors.
(Note that the choice of $\mu^2$ in Figure~\ref{C2} is only for
the purpose of comparison with the other methods, which introduce
charm at $m_c^2$.
In the final calculations the scale $\mu^2 = 4 m_c^2$ will be used
for the FFNS, which is also the value used in Ref.\cite{GRV}.)

At small $x$ the effect of the mass-corrected quark coefficient
functions on $F_2^c$ turns out to be negligible, and only slight
at larger $x$.
As $Q^2$ becomes large, one can see in Figure \ref{C2} how the
VFNS and interpolating schemes converge.
Even at small $Q^2$ the difference between these is not large.  
On the other hand, while the FFNS provides a good approximation to
the interpolating scheme for $Q^2 \alt 30$ GeV$^2$, it dramatically
overestimates the full result at larger $Q^2$, especially at large $x$.
Since this is the region where most of the relevant EMC data which
we analyse lie, clearly we need to use the full interpolating scheme
in order to draw reliable conclusions from our analysis.

Another source of uncertainty in the calculation of $F_2^c$
comes from the gluon distribution at large $x$, which at present
is not very well constrained.
To cover the full range of allowed gluon distributions, we use the
maximum and minimum gluon distributions from the MRST \cite{MRST} 
parameterizations in addition to that of GRV \cite{GRV}.
The resulting $F_2^c$ for the different next-to-leading order
glue is shown in Figure~\ref{C3}, where evolution was again
performed as described in the massless evolution section
The data at small $x$ in Figure~\ref{C3} are taken from the ZEUS
Collaboration \cite{ZEUS}, while the large-$x$ data are from the
earlier EMC experiment \cite{EMC}.
At small $x$ all of the parameterizations fit the data very well.
At large $x$ the maximum-gluon MRST and GRV fits also provide
good descriptions of the data, with the exception of the two
points at $Q^2=60$ GeV$^2$.
The last data point at $x=0.44$ is not shown in the fits of
Refs.\cite{MRST,GRV}, even though this point appears to confirm
the trend indicated by the $x=0.24$ point to lie somewhat above
the perturbative QCD calculation.
Neglecting the large-$x$, $Q^2=60$ GeV$^2$ points, one would
conclude that perturbative QCD fits the $F_2^c$ data very well,
without any need for additional non-perturbative contributions
\cite{MRST,GRV}.
On the other hand, taking these points seriously has led several
authors \cite{HM,HSV} to conclude that the large-$x$ EMC data
provide evidence for an intrinsic charm component of the nucleon.

In the next Section we shall study the large-$x$ EMC data
more carefully, with the aim of ascertaining whether these
can be understood perturbatively, or whether they
can indeed be interpreted as suggesting that a perturbative QCD
treatment alone is incomplete.

\section{Non-Perturbative Charm}
 
The apparent discrepancy between some of the large-$x$ $F_2^c$
data and predictions based solely on perturbative QCD has
prompted several authors \cite{BROD,HM,HSV} to take seriously
the possibility that an additional, non-perturbative, component
of $F_2^c$ may be necessary to account for the data over the
full range of $x$ and $Q^2$.
In this Section we discuss how non-perturbative charm may
affect $F_2^c$, particularly at large $x$, and how the
intrinsic contributions can be included on the same footing
as the perturbative effects.

In earlier analyses \cite{HM,HSV} intrinsic charm distributions
have simply been added to the perturbatively generated $F_2^c$,
\begin{equation}
\label{e5}
F_2^c(x,Q^2) = F_2^{c\ ({\rm pert})}(x,Q^2)
	     + F_2^{c\ ({\rm IC})}(x,Q^2),
\end{equation}
where the perturbative contribution, $F_2^{c\ ({\rm pert})}$, is given
by Eq.(\ref{pgf}) (with higher order corrections),
while the intrinsic charm contribution, $F_2^{c\ ({\rm IC})}$,
in its simplest form is:
\begin{equation}
\label{icsimple}
F_2^{c\ ({\rm IC})} = e_c^2\ x (c^{\rm IC} + \bar c^{\rm IC}).
\end{equation}
In practice, ${\cal O}(\alpha_s)$ contributions to
Eq.(\ref{icsimple}) are fully implemented in this
analysis (the relevant expressions are given in
Ref.\cite{HM}).

Within the interpolating scheme of Section IV, the most natural
way to implement intrinsic charm in $F_2^c$ is to modify the
boundary condition for the charm quark distribution.
Instead of $c(x,\mu^2) = \bar c(x,\mu^2) = 0$, one now has
non-zero distributions at the scale $\mu^2=m_c^2$.
The physical reasoning for this is that if there are
non-perturbative processes producing charm in the nucleon,
this charm can be resolved (brought on its mass shell)
only when the system has sufficient energy.
In the $\overline{\rm MS}$ scheme, the scale at which the
number of flavors changes from 3 to 4 is $\mu^2=m_c^2$,
so that regardless of the dynamical origin of the charm,
there will be enough energy in the system to open a new
active flavor channel for $Q^2 > m_c^2$.
With this in mind, we next discuss several non-perturbative
models which attempt to describe the generation of intrinsic
charm in the nucleon.

\subsection{Five-Quark Component of the Nucleon (IC1)}
 
Based on the initial observation \cite{HADR} that the charm
production cross section in hadronic collisions was larger
than that predicted in leading order perturbative QCD,
Brodsky et al. \cite{BROD} suggested that the discrepancy
could be resolved by introducing an intrinsic, non-perturbative,
charm component in the nucleon wave function.
In this model, which we shall refer to as ``IC1'', the nucleon
is assumed to contain, in addition to the lowest energy three-quark
Fock state, a more complicated, five-quark configuration on the
light-cone:
\begin{eqnarray}
| p \rangle
&=& c_0 | uud \rangle + c_1 |uudc\bar c \rangle ,
\end{eqnarray}
where $c_0^2\ (c_1^2)$ is the three (five)-quark probability.
In order to explain the original data \cite{HADR}, the
normalization of the latter was chosen to be 1\% \cite{BROD}.
Assuming the five-quark wave function to be inversely proportional
to the light-cone energy difference between the nucleon ground
state and the five-quark excited state, one finds that the
$x$-dependence of the $c$ quark distribution is given by
\cite{BROD}:
\begin{eqnarray}
\label{delic}
c^{\rm IC1}(x)  
&=& 6 x^2 \left( (1-x) (1 + 10x + x^2) - 6x (1+x) \log 1/x \right).
\end{eqnarray}  
The anticharm distribution has the same shape as 
the charm distribution in this model,
$\overline c^{\rm IC1}(x) = c^{\rm IC1}(x)$.

Because the intrinsic charm in this model is assumed to be
generated through $gg \rightarrow c\bar c$ processes, with each
gluon originating from different valence quarks, the $c\bar c$
probability scales like $\alpha_s^2(m_c^2)/m_c^2$ relative to
the perturbative component \cite{HSV,ICHT}.
This contribution was therefore interpreted in Refs.\cite{HSV,ICHT}
as a higher-twist effect.
Our philosophy here is that since this is generated non-perturbatively,
the resulting non-perturbative intrinsic charm distribution calculated
at $m_c^2$ should be evolved as leading twist, on the same footing as
the perturbative contribution.
Since there is only one kind of charm quark, irrespective of its
origin, QCD corrections should affect both the perturbative and
non-perturbative distributions identically.
%

\subsection{Meson Cloud Model (IC2)}

An alternative to the five-quark intrinsic charm model was
considered in Refs.\cite{NAVARRA,PAIVA,MT}, in which the
charm sea was assumed to arise from the quantum fluctuation of
the nucleon to a virtual $\bar D^0 + \Lambda^+_c$ configuration.  
In the following we shall refer to this model as ``IC2''.
The nucleon charm radius \cite{NAVARRA} and the charm quark
distribution \cite{PAIVA} were both estimated in this framework.
Furthermore, the effects of hard charm distributions on large-$x$
HERA cross sections, and in particular on the so-called HERA anomaly
\cite{HERANOM,GV}, were studied in Ref.\cite{MT}.

The meson cloud model for the long-range structure of the nucleon
has been used extensively to describe various flavor symmetry
breaking phenomena observed in deep-inelastic scattering and
related experiments.
It offers a natural explanation of the $\overline d$ excess in
the proton over $\overline u$ \cite{EXCESS,E866} in terms of a
pion cloud, which itself is a necessary ingredient of the nucleon
by chiral symmetry.
It also provides an intuitive framework to study the strangeness
content of the nucleon, through the presence of the kaon cloud
\cite{STRANGE}.
Whether the same philosophy can be justified for a cloud of heavy
charm mesons and baryons around the nucleon is rather more
questionable given the large mass of the fluctuation.
Nevertheless, to a crude approximation, one may take the meson
cloud framework as indicative of the possible order of magnitude
and shape of the non-perturbative charm distribution.
Furthermore, a natural prediction of this model is that the 
$c$ and $\overline c$ distributions are not symmetric.

In the meson cloud model, the distribution of charm and anticharm
quarks in the nucleon at some low hadronic scale can be approximated
by \cite{MT}:
\begin{mathletters}
\label{mcm_final}
\begin{eqnarray}
c^{\rm IC2}(x) &\approx& {3 \over 2} f_{\Lambda_c/N}(3x/2),	\\
\overline c^{\rm IC2}(x) &\approx& f_{\bar D/N}(x),
\end{eqnarray}
\end{mathletters}%
where 
\begin{eqnarray}
\label{fz}
f_{\bar D/N}(x)
&=& { 1 \over 16 \pi^2 }
\int_0^\infty dk^2_\perp
{ g^2(k_\perp^2,x) \over x (1-x) (s - M^2)^2 }
\left( { k_\perp^2 + [M_{\Lambda_c} - (1-x) M]^2 \over 1-x } \right)
\end{eqnarray}
is the light-cone distribution of $\bar D^0$ mesons in the nucleon,
and $f_{\Lambda_c/N}(x) = f_{\bar D/N}(1-x)$ is the corresponding
distribution of $\Lambda_c^+$ baryons.
In Eq.(\ref{fz}) the function $g$ describes the extended nature
of the $\bar D\Lambda_c N$ vertex, with the momentum dependence
parameterized by\
$g^2(k_\perp^2,x)\
=\ g_0^2\ (\Lambda^2 + M^2)/(\Lambda^2 + s)$,
where $s$ is the $\bar D\Lambda_c$ center of mass energy squared
and $g_0$ the $\bar D \Lambda_c N$ coupling constant at the pole,
$s = M^2$.
As a first approximation, one might take $g_0$ to be of the
same order of magnitude as the $\pi NN$ coupling constant.
In Ref.\cite{NN} this coupling constant was estimated within
a QCD sum rule calculation.

\subsection{Intrinsic Charm Distributions}

The $c$ and $\overline c$ distributions in the intrinsic
charm models IC1 and IC2 are shown in Figure~\ref{C4},
each normalized to a common value of 1\%.
For the IC2 model this corresponds to a cutoff
$\Lambda \approx 2.2$~GeV (for a probability of 0.5\%, for
example, one would need $\Lambda \approx 1.7$~GeV).
Quite interestingly, the shapes of the $c$ quark distributions
are quite similar in the two models, with $xc$ peaking at around
$x \sim 0.3$.
However, because the IC2 model gives a significantly harder
$\overline c$ distribution, while IC1 implies that $c$ and
$\bar c$ are equal, the resulting structure function, $F_2^c$,
will be somewhat harder in the IC2 model.
For comparison, a typical (soft) perturbatively-generated
charm distribution is also shown in Figure~\ref{C4},
evaluated from the MRST parameterization \cite{MRST}
(with the maximal gluon) at $Q^2=5$ GeV$^2$.

The effects of the modified boundary conditions incorporating
non-zero intrinsic charm distributions are shown in Figure~\ref{C5}
for the GRV parameterization \cite{GRV} and for 1\% intrinsic
charm normalizations, at $Q^2 = 25$, 45 and 60 GeV$^2$.
The data at 60 GeV$^2$ are well fitted with a 1\% IC1 component,
although with the IC2 model one slightly overestimates the $x=0.44$
data point, due to its very hard $\bar c$ distribution.
At lower $Q^2$ values, however, the addition of a 1\% intrinsic
charm component, from either model, overestimates the large-$x$
points.
This finding is essentially independent of the parton distribution
functions employed, as Figure~\ref{C6} illustrates for the MRST
distributions \cite{MRST}.
{}From this one can conclude that with a 1\% intrinsic charm component
one cannot simultaneously resolve the large-$x$ discrepancy for the
large-$Q^2$ data, and maintain a satisfactory fit to the data at
lower $Q^2$.
One could imagine introducing a $Q^2$-dependent normalization for
the intrinsic charm to improve the fits to the 25 and 45 GeV$^2$ data,
although this solution would not be well grounded theoretically.

To compare with the procedure for incorporating intrinsic charm
adopted in the earlier analysis in Ref.\cite{HM}, we show in
Figure~\ref{C7} the $F_2^c$ obtained from the FFNS through
Eq.(\ref{e5}), and the ${\cal O}(\alpha_s)$ corrections to
the intrinsic charm component, $F_2^{c\ ({\rm IC})}$ ---
see Eq.(57) of Ref.\cite{HM}.
(Note that although the FFNS curves appear to lie slightly
below the data for $Q^2=25$ and 45 GeV$^2$, inclusion of
${\cal O}(\alpha_s^2)$ corrections leads to a slight
improvement for the lower $Q^2$ data, without much effect
on the data at 60 GeV$^2$ \cite{GRV}.)
The results are qualitatively similar to those obtained from
the full interpolating scheme, namely the data at different $Q^2$
seem to require different functional forms and normalizations for
the intrinsic charm.

By varying the amount of intrinsic charm that the data can
accommodate, one can attempt to improve the overall fit at
all values of $Q^2$.
Using the minimum gluon fit of the MRST parameterization one
can determine the maximum intrinsic charm allowed by the data
for both the IC1 and IC2 models.
Slightly better fits can be obtained with a 0.75\% IC1 charm
distribution, and with a 0.4\% IC2 distribution, as indicated
in Figure~\ref{C8} for $Q^2=45$ and 60 GeV$^2$.
(The quality of the fit at $Q^2=25$ GeV$^2$ is similar
to that for the 45 GeV$^2$ data.)
However, even given the different shapes of the charm and
anticharm distributions in the IC1 and IC2 models, it is still
quite difficult to obtain a quality fit to data at all $Q^2$
values.
For the Hoffman and Moore procedure \cite{HM} with the FFNS,
one finds similar difficulties in reconciling the $Q^2=60$ GeV$^2$
data with those at lower $Q^2$, even as a function of the intrinsic
charm model, and normalization (c.f. Figure~\ref{C9}), although there
may be a slight preference for IC2 with 0.4\% normalization.

In view of the difficulties in obtaining a simultaneous fit to
the $F_2^c$ data at all measured values of $x$ and $Q^2$ with
either the perturbative-only, or the intrinsic charm scenarios,
it seems imperative that the data, particularly those at
large $x$ and $Q^2$, be confirmed by independent measurements.

\section{Conclusion}

Unraveling the rich structure of the nucleon sea is an ongoing
endeavor which has taught us a number of important and sometimes
surprising lessons in recent years.
Far from being a homogeneous background in which the valence
quarks reside, the sea has proved to exhibit considerable
intrinsic structure of its own.
Perhaps the most conspicuous structure is that in the light quark
sea of the proton, where a number of experiments have now confirmed
beyond any doubt a significant excess of $\bar d$ over $\bar u$
quarks \cite{EXCESS}.
This in turn has created an environment in which the importance
of {\em non-perturbative} effects in the nucleon is appreciated
to a far greater extent, even when discussing structure at
deep-inelastic energy scales \cite{E866}.

More speculative, and less constrained experimentally, are
suggestions that the proton sea for heavier flavors might
also exhibit characteristics which cannot be attributed
to perturbative QCD mechanisms alone \cite{MT,STRANGE}.
A prime example would be the presence of asymmetric sea quark
and antiquark distributions, which have been searched for in
the strange sector in both deep-inelastic neutrino and
antineutrino scattering, as well as in electromagnetic form
factors at low energies \cite{SFF}.
Indeed, there is no symmetry principle in QCD which would
prevent sea quarks and antiquarks having different momentum
distributions, just as there is no symmetry requiring the
$\bar d$ and $\bar u$ sea to be equivalent.

More challenging still is to identify possible non-perturbative
effects in the charm sector, especially since the magnitude of
the total charm component of non-charm hadrons is tiny.
Nevertheless, a number of pioneering studies \cite{BROD,HM,HSV}
of various charm production reactions have left us with lingering
doubts as to whether perturbative QCD is the full story behind
these processes.
On the other hand, the evidence doesn't appear to be conclusive
enough to warrant introduction of intrinsic charm distributions
in global data parameterizations \cite{MRST,GRV,CTEQ5}, which
to date have generated the charm structure function purely
perturbatively.

The present analysis has been an attempt to resolve the issue
of whether the charm electroproduction data do indeed support
the existence of non-perturbative components of the nucleon
wave function, or whether they can be understood within
conventional perturbative dynamics.
To this effect we have used the latest available technology
to describe charm production over the entire range of $x$
and $Q^2$ accessible to experiment.
Our approach consistently interpolates between the two
asymptotic regions of massless evolution at large $Q^2$
and the photon--gluon fusion process at low $Q^2$, and
includes quark and target mass effects and corrections
for mass thresholds.

To a certain extent our findings confirm the existing state of affairs,
in which some of the data show no evidence at all for intrinsic
charm, while other data cannot be fitted without additional
non-perturbative input.
Even within the rather different models of intrinsic charm considered
here, with varying normalizations, it seems difficult to simultaneously
fit the entire data set in terms of a single intrinsic charm scenario
--- although there may be a slight preference for intrinsic charm in
model IC2 at a level of about 0.4\%.
The clearest conclusion that one can draw from this is that more
quality data are urgently needed to settle the issue.
In particular, while the small-$x$ domain seems to be relatively
well under control, the large-$x$ region, where cross sections are
small and measurements more difficult, must be the focus of future
experimental effort if the non-perturbative structure of the nucleon
sea is to be resolved.

\acknowledgements

W.M. and F.M.S. would like to acknowledge the support of the Special
Research Centre for the Subatomic Structure of Matter at the University
of Adelaide during the initial stages of this work.
F.M.S. would like to thank S.J. Brodsky, S. Kretzer and F.S. Navarra
for discussions.
This work was supported by the Australian Research Council and by
FAPESP (Brazil).


\begin{figure}
\hspace*{2cm}
\epsfig{figure=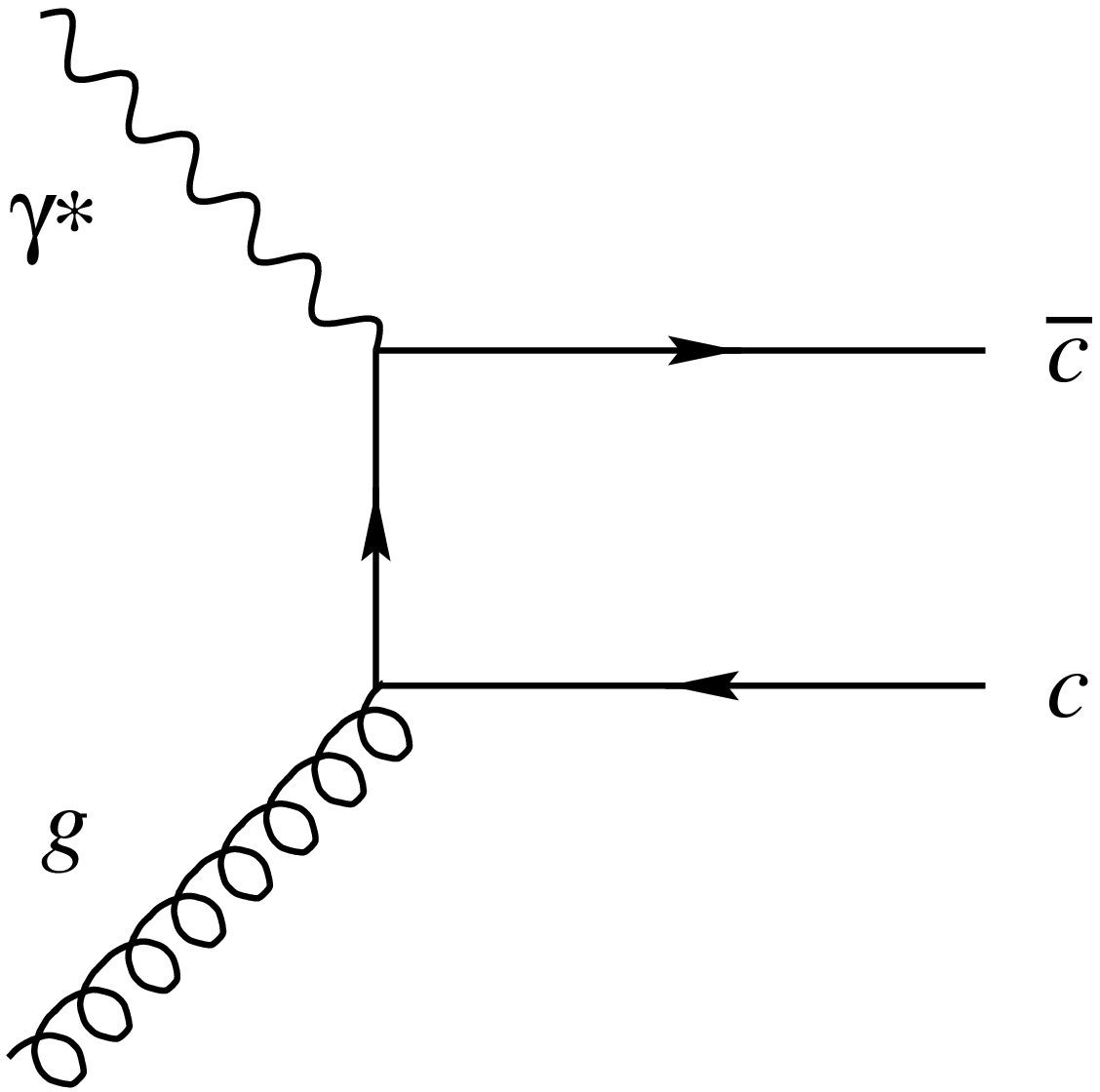,height=8cm}
\vspace*{1cm}
\caption{Photon--gluon fusion process at leading order in $\alpha_s$.}
\label{C1}
\end{figure}

\begin{figure}
\centering{
\begin{picture}(120,200)(180,0)
\epsfig{figure=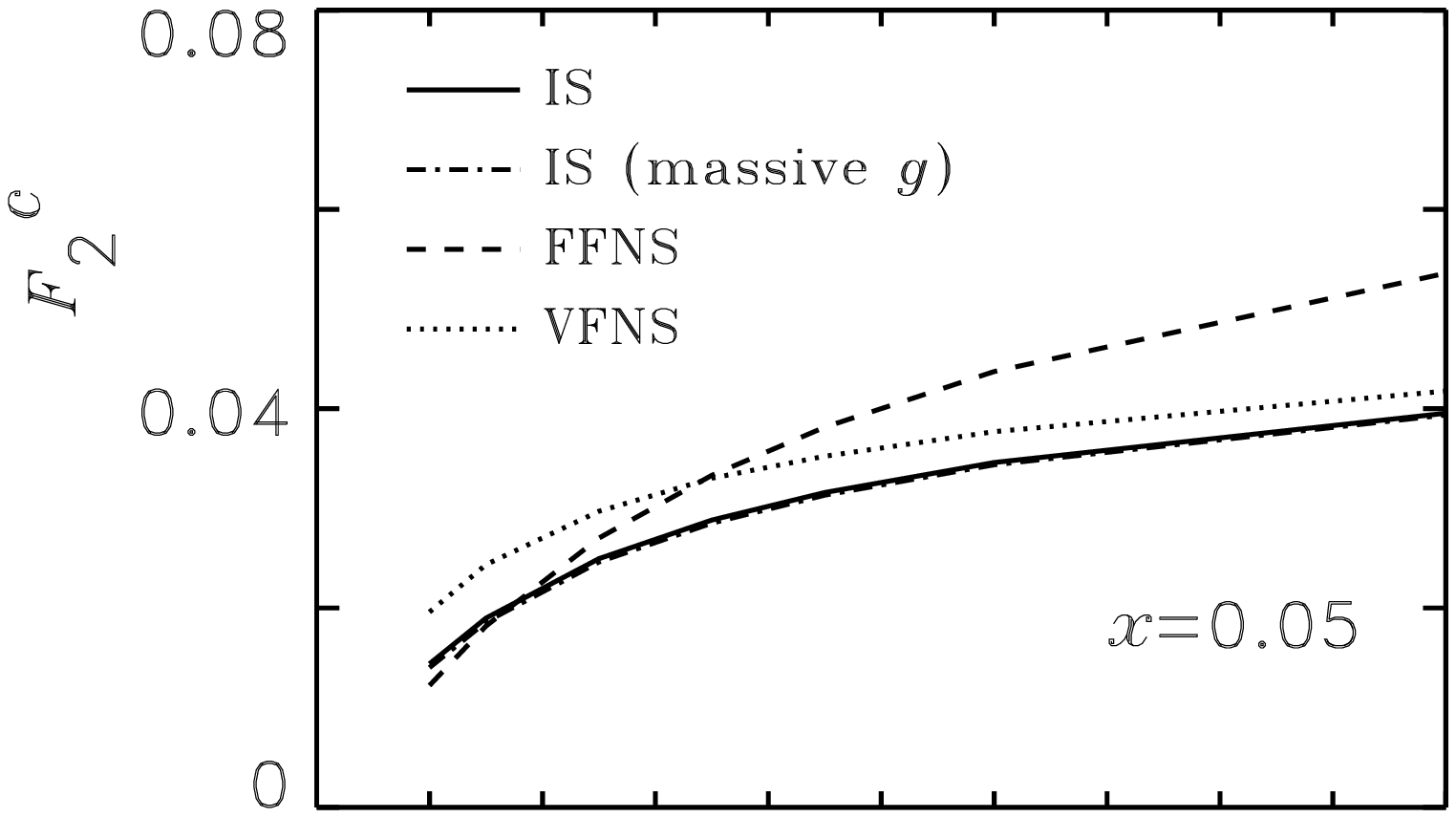,height=8cm}
\put(-319,-160){\epsfig{figure=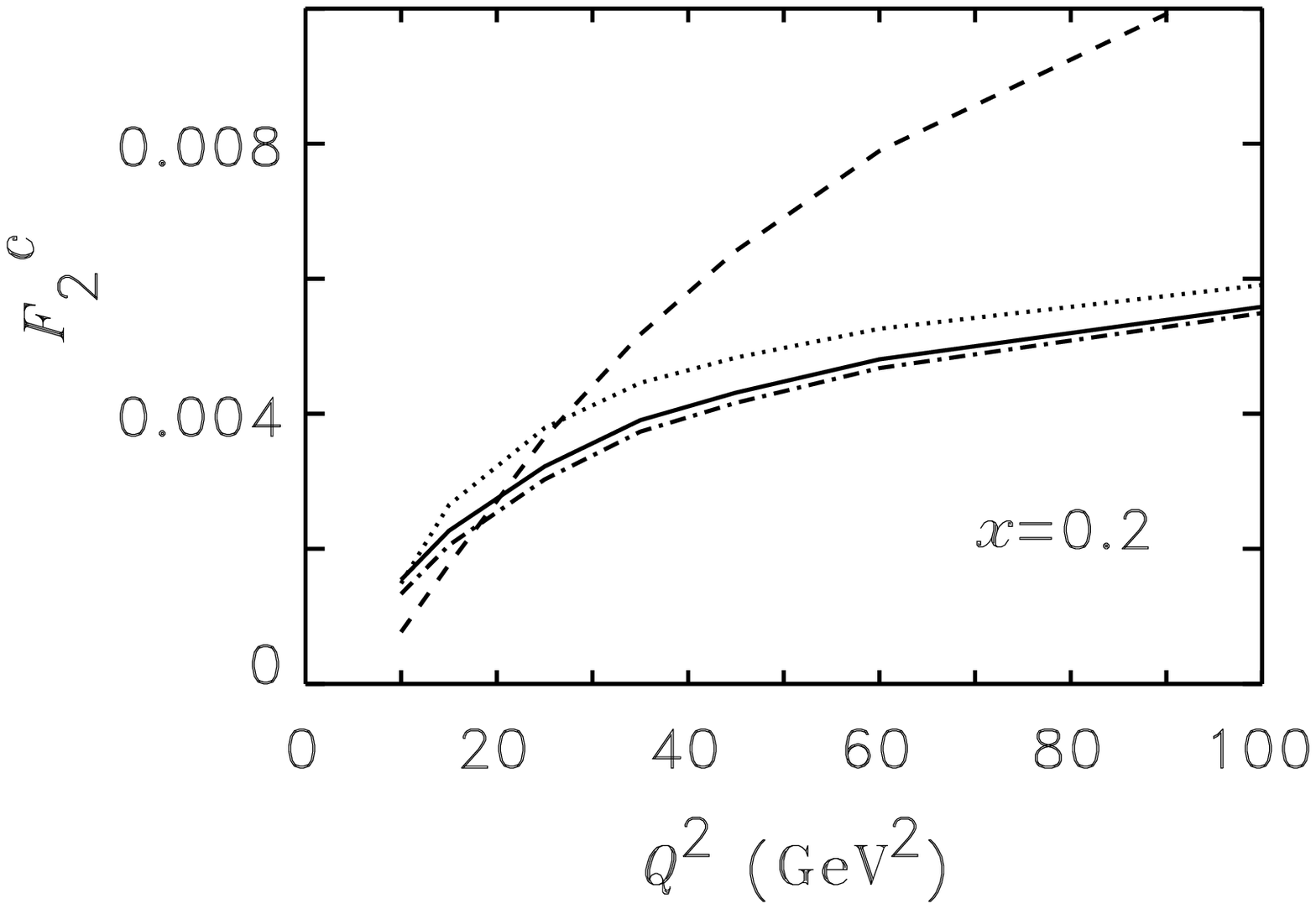,height=8cm}}
\end{picture}}
\vspace*{7cm}
\caption{Charm structure function as a function of $Q^2$
	for $x=0.05$ and 0.2, evaluated according to the
	various schemes discussed in the text.
	``IS'' refers to the interpolating scheme of
	Eq.(\protect\ref{interp}), while ``IS (massive $g$)''
	denotes the interpolating scheme but with massless
	quark coefficient functions.
	The GRV parameterization \protect\cite{GRV} for the
	gluon is used.}
\label{C2}
\end{figure}

\begin{figure}
\epsfig{figure=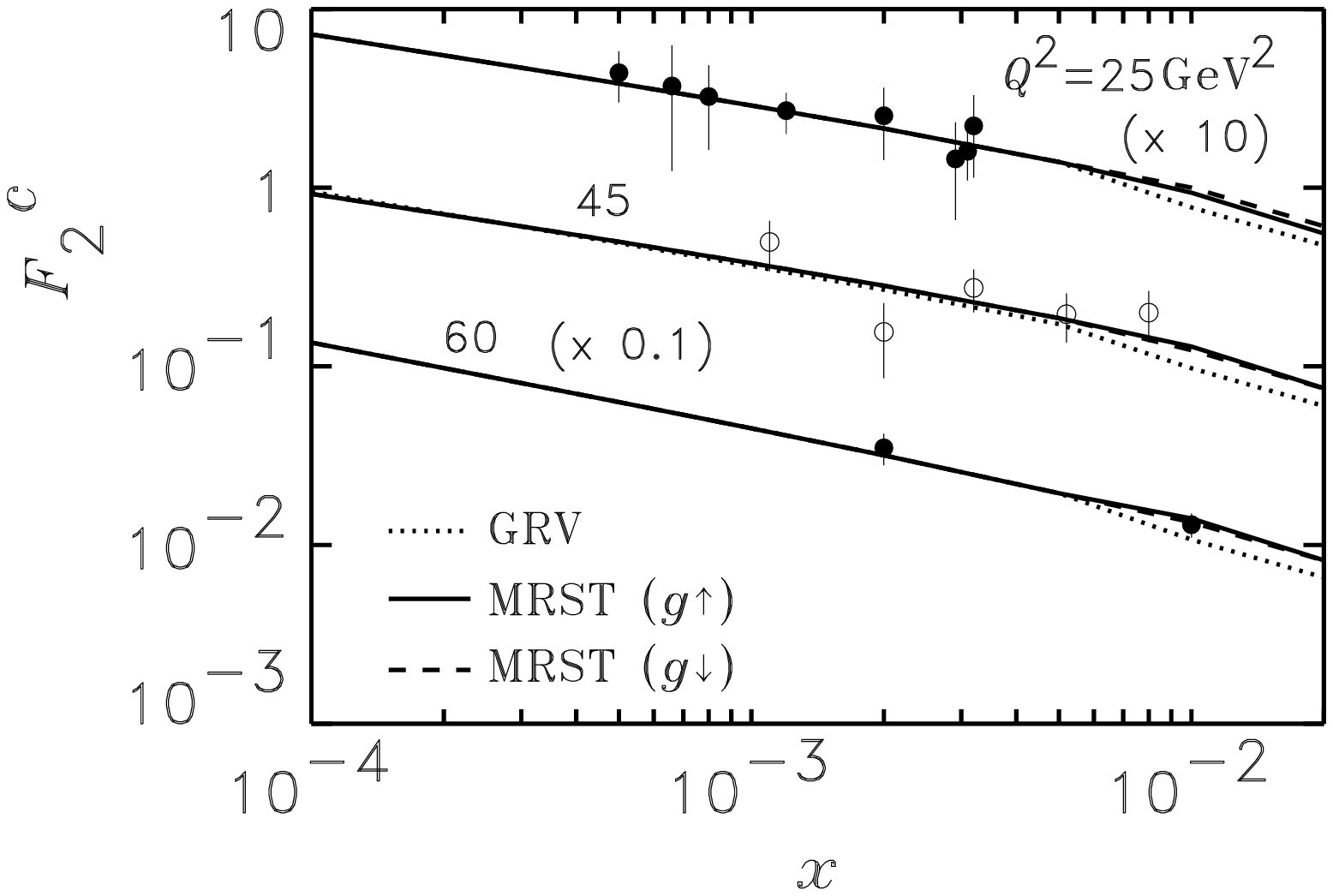,height=8cm}
\epsfig{figure=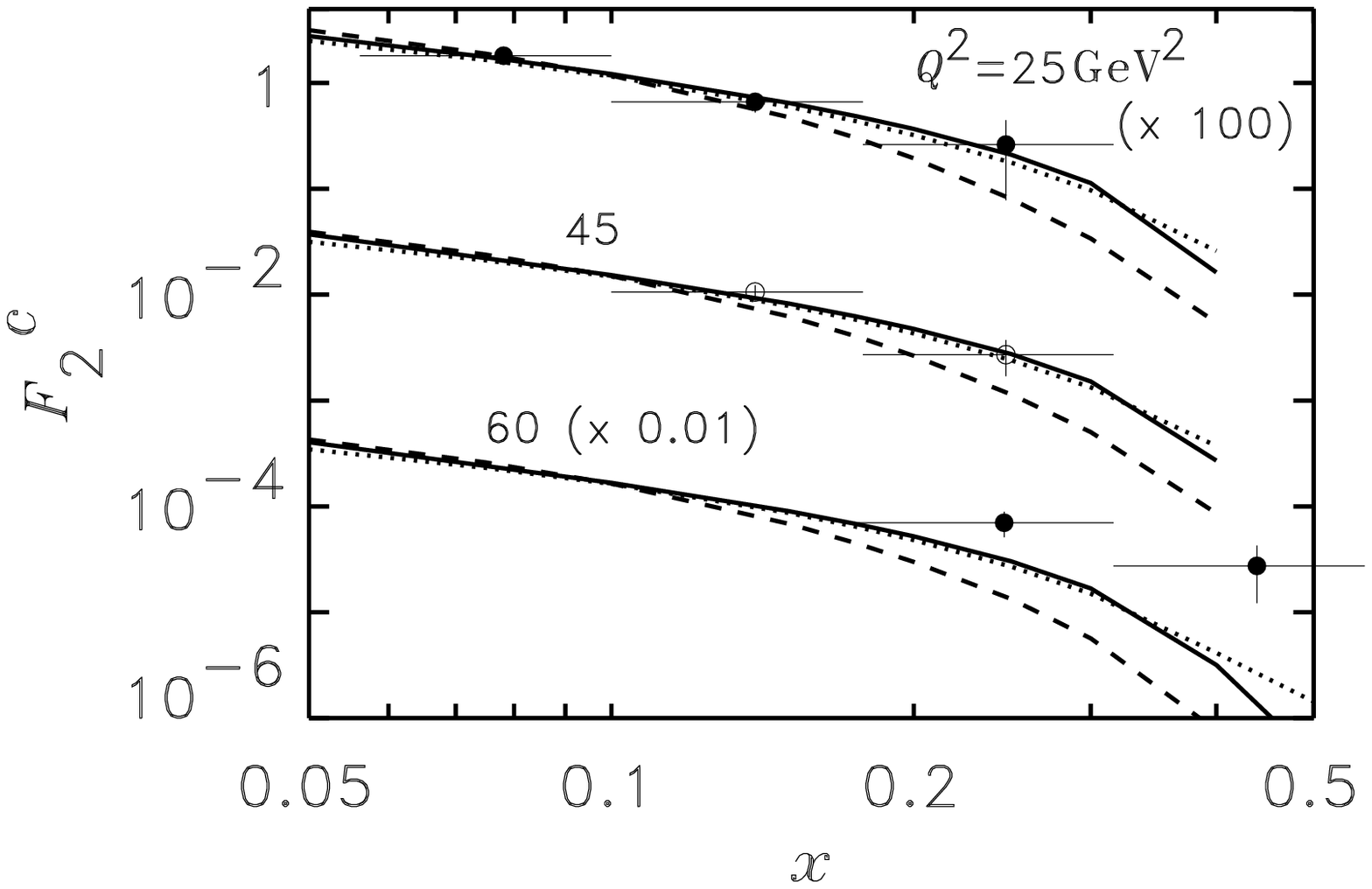,height=8cm}
\caption{Charm structure function calculated within the
	interpolating scheme for different gluon distributions,
	GRV \protect\cite{GRV} (dotted) and MRST \protect\cite{MRST}
	with minimum (dashed) and maximum (solid) gluons.
	The data at small $x$ are from the ZEUS Collaboration
	\protect\cite{ZEUS}, while the large-$x$ data are from
	the EMC \protect\cite{EMC}.
	For clarity the small-$x$ curves have been scaled by a
	factor 10 (0.1) for $Q^2=25$ (60) GeV$^2$, and the large-$x$
	curves by a factor 100 (0.01).}
\label{C3}
\end{figure}

\begin{figure}
\epsfig{figure=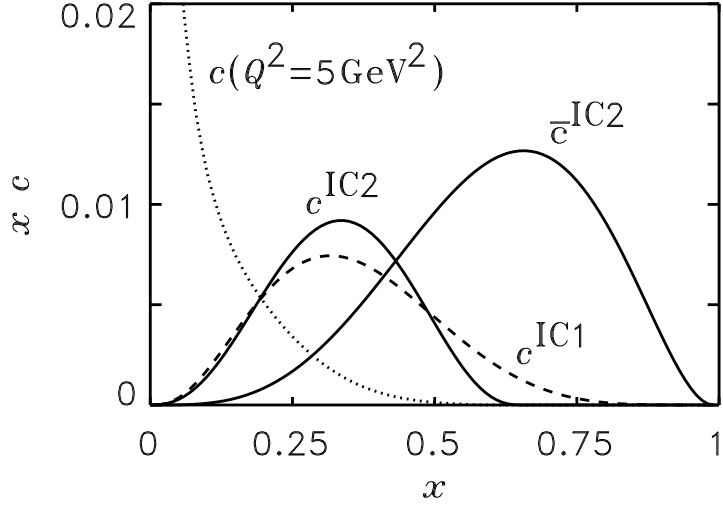,height=8cm}
\caption{Charm quark distributions from the intrinsic charm models
	IC1 \protect\cite{MT} (solid) and IC2 \protect\cite{BROD}
	(dashed), both normalized to 1\%, and from the MRST
	parameterization \protect\cite{MRST} (with maximal gluon)
	at $Q^2=5$ GeV$^2$ (dotted).}
\label{C4}
\end{figure}

\begin{figure}
\centering{
\begin{picture}(120,150)(180,0)
\epsfig{figure=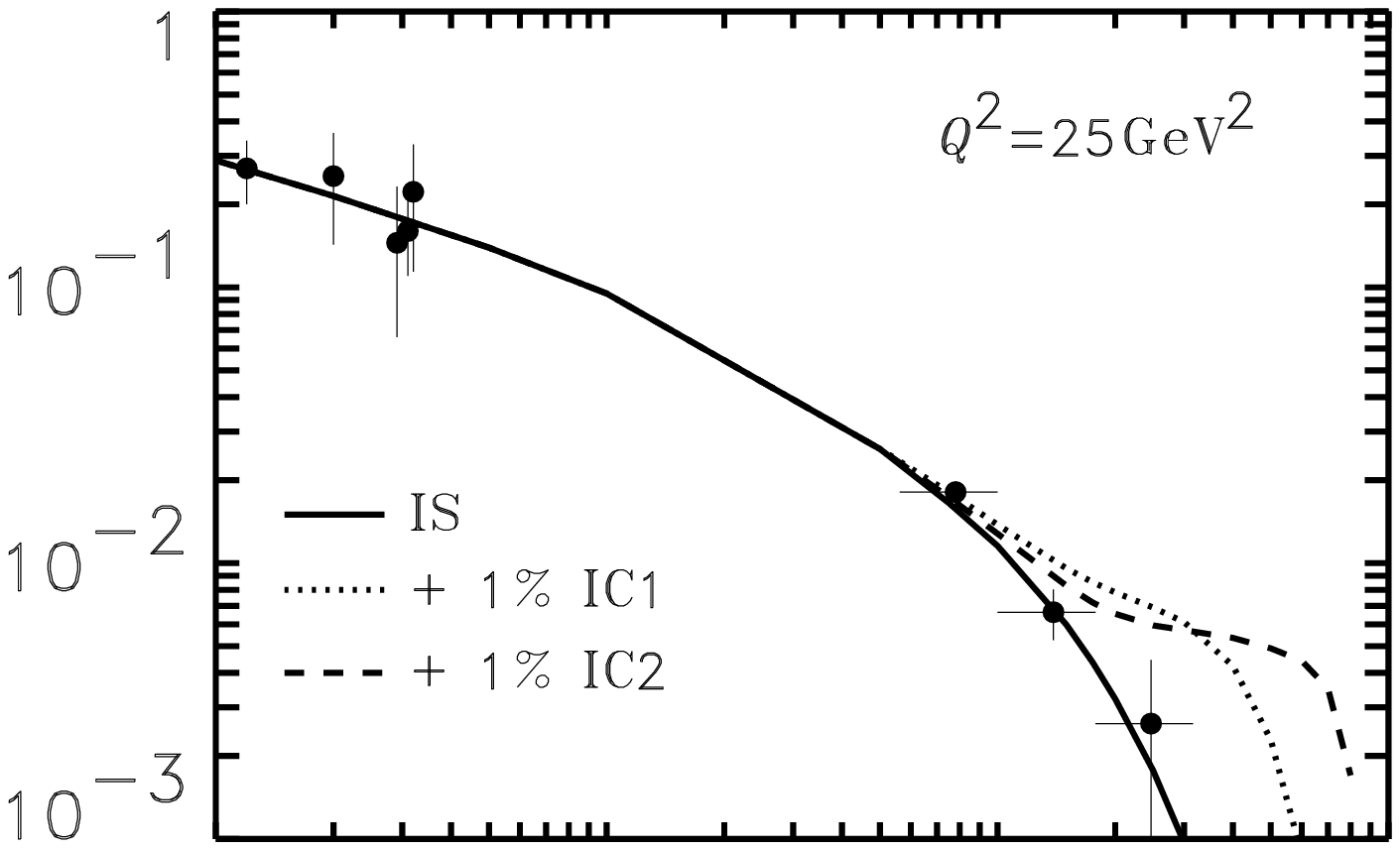,height=7cm}
\put(-279,-135){\epsfig{figure=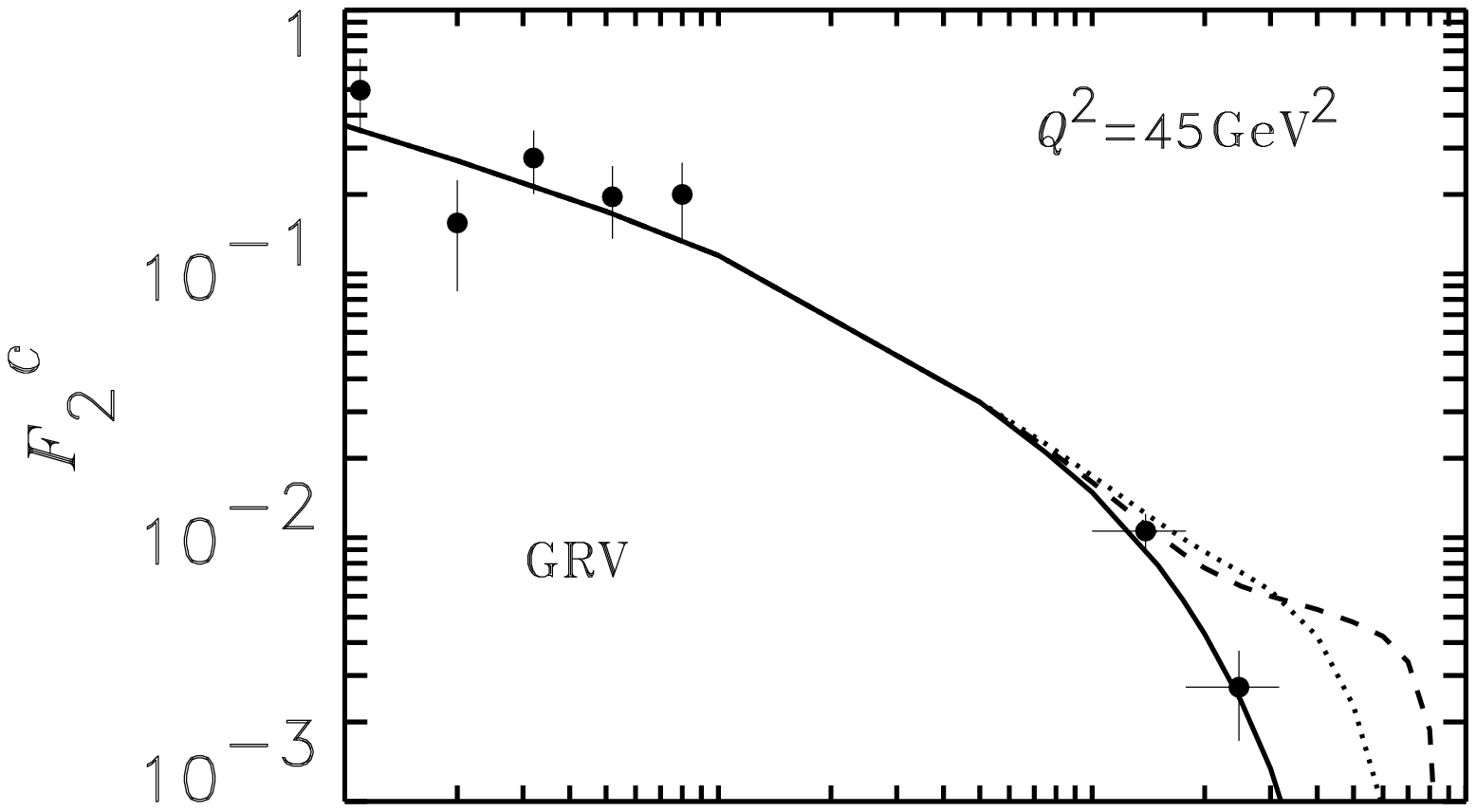,height=7cm}}
\put(-279,-270){\epsfig{figure=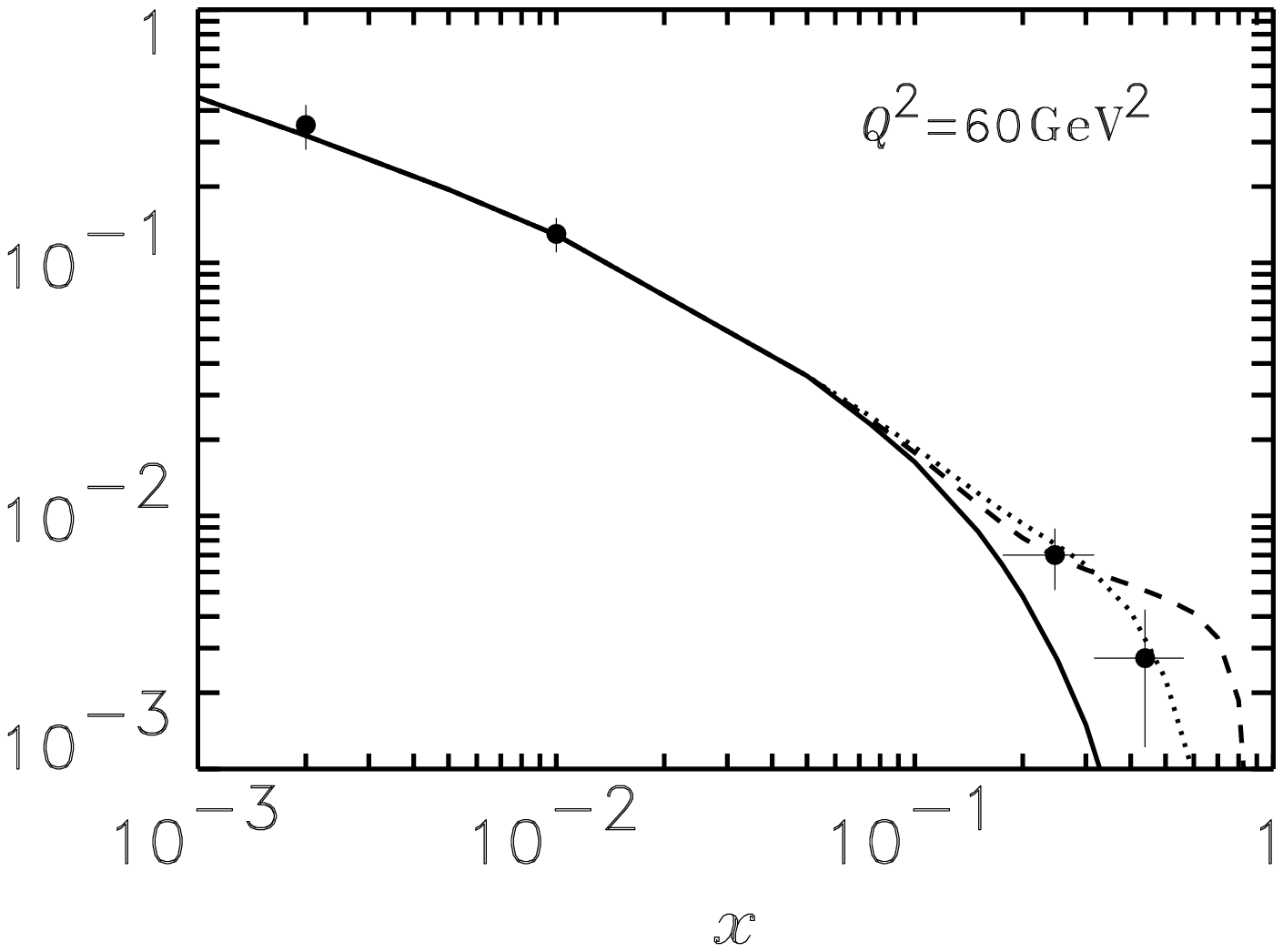,height=7cm}}
\end{picture}}
\vspace*{10cm}
\caption{The charm structure function calculated through the
	interpolating scheme (``IS'') in Eq.(\protect\ref{interp})
	for intrinsic charm distributions from models IC1 and IC2,
	normalized to 1\%.
	The GRV parameterizations \protect\cite{GRV} for the gluon
	and light quark densities are used.}
\label{C5}
\end{figure}

\begin{figure}
\centering{
\begin{picture}(120,150)(180,0)
\epsfig{figure=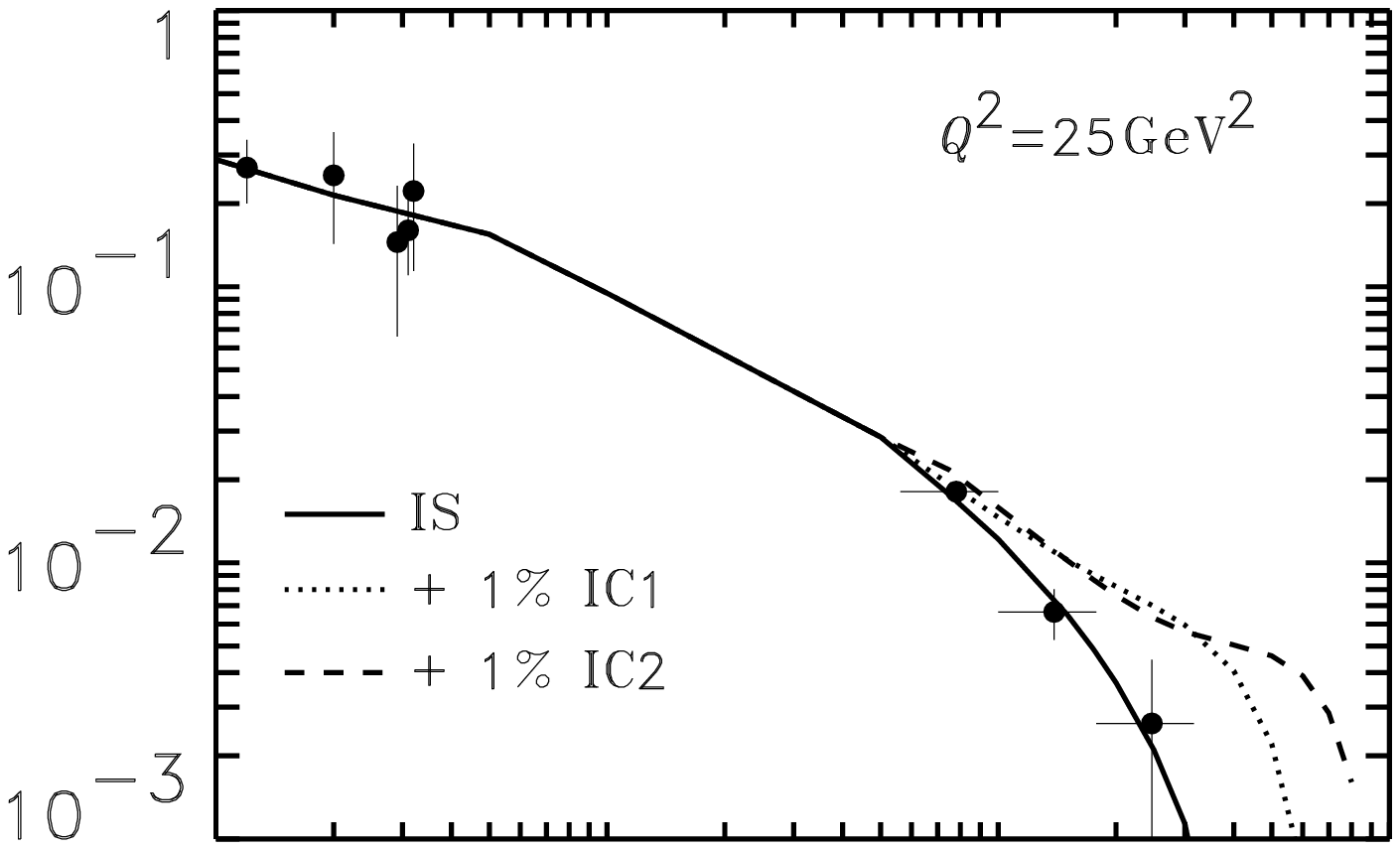,height=7cm}
\put(-279,-135){\epsfig{figure=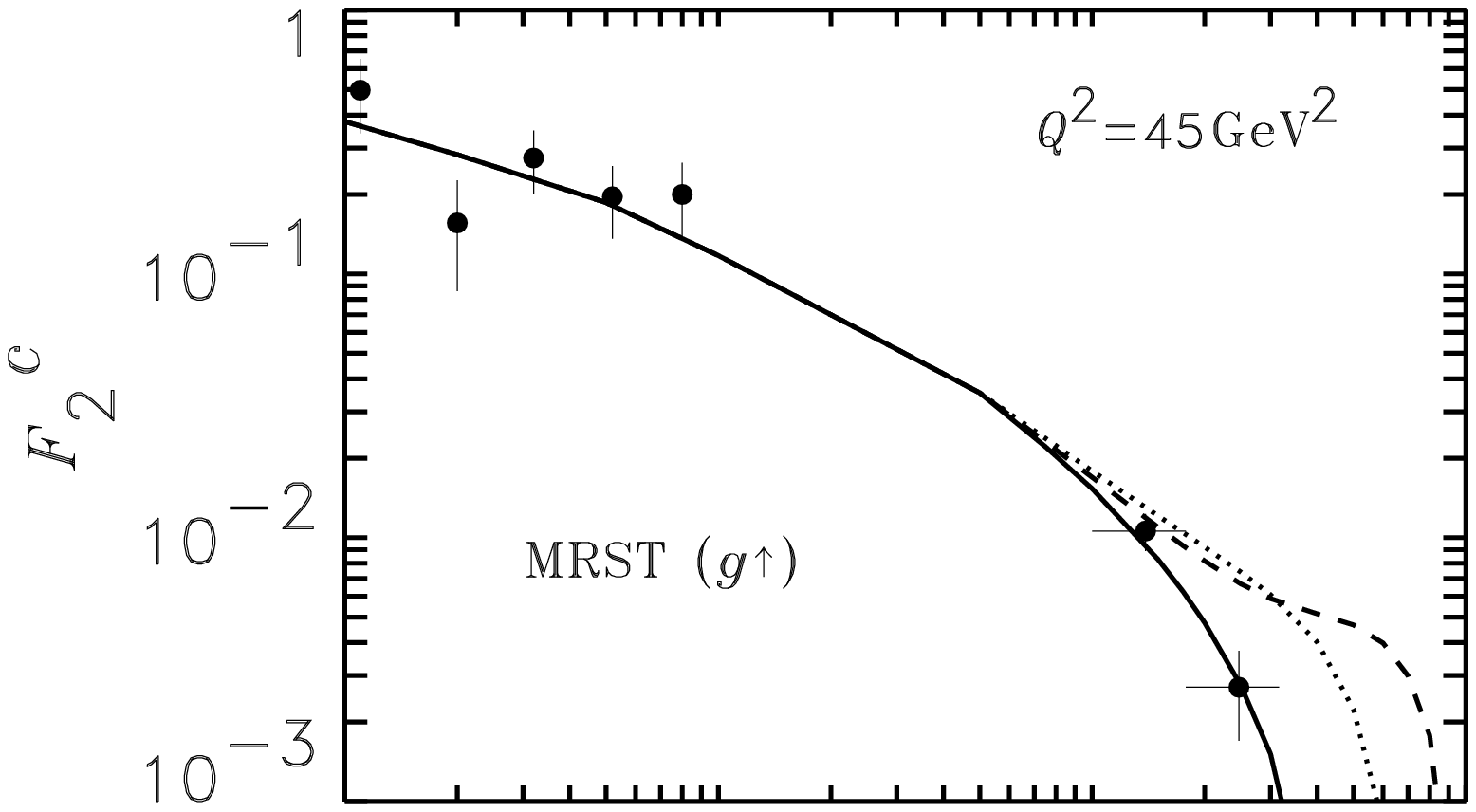,height=7cm}}
\put(-279,-270){\epsfig{figure=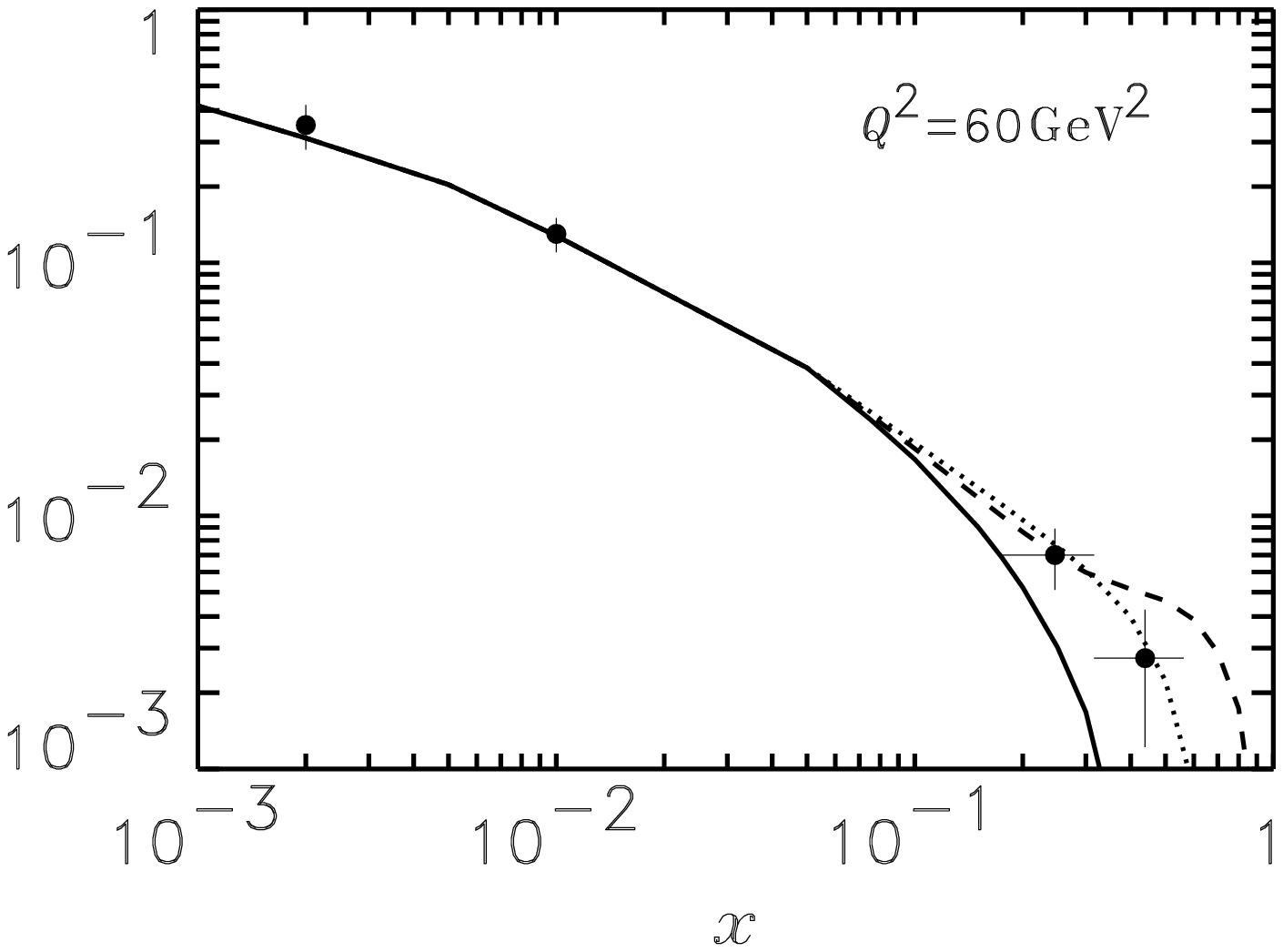,height=7cm}}
\end{picture}}
\vspace*{10cm}
\caption{As in Figure~\protect\ref{C5}, but with the MRST
	parameterization \protect\cite{MRST} with the maximum gluon.}
\label{C6}
\end{figure}

\begin{figure}
\centering{
\begin{picture}(120,150)(180,0)
\epsfig{figure=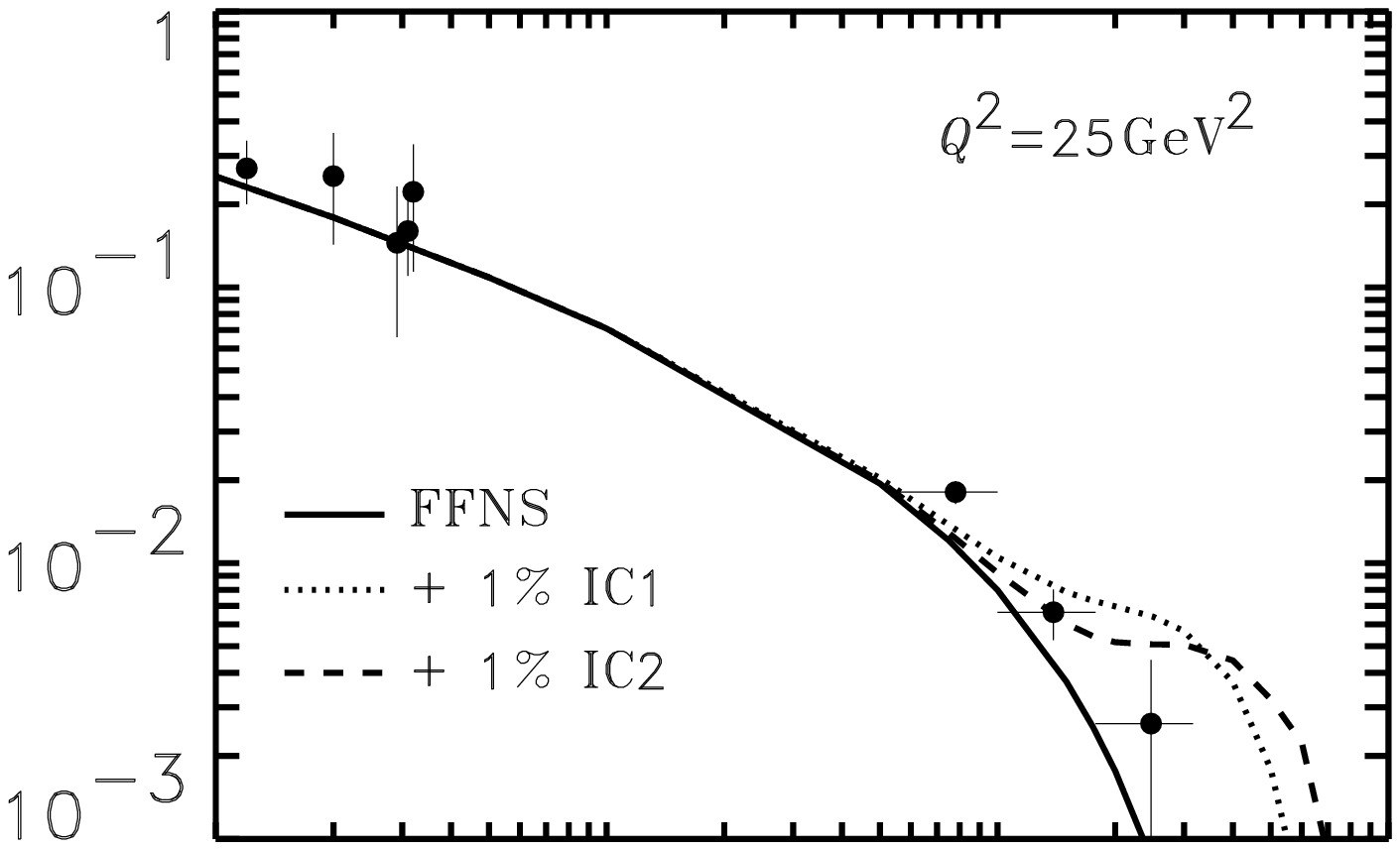,height=7cm}
\put(-279,-135){\epsfig{figure=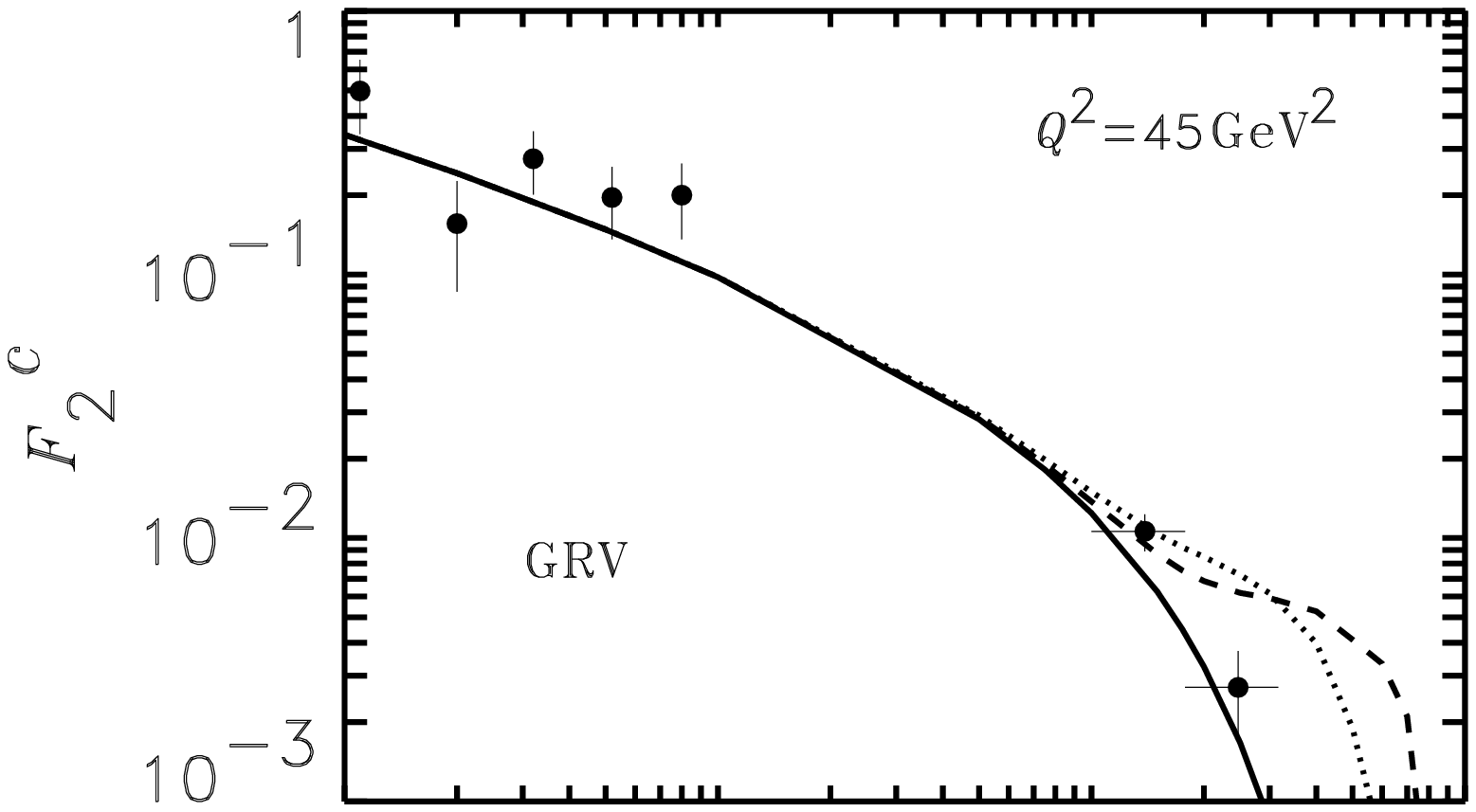,height=7cm}}
\put(-279,-270){\epsfig{figure=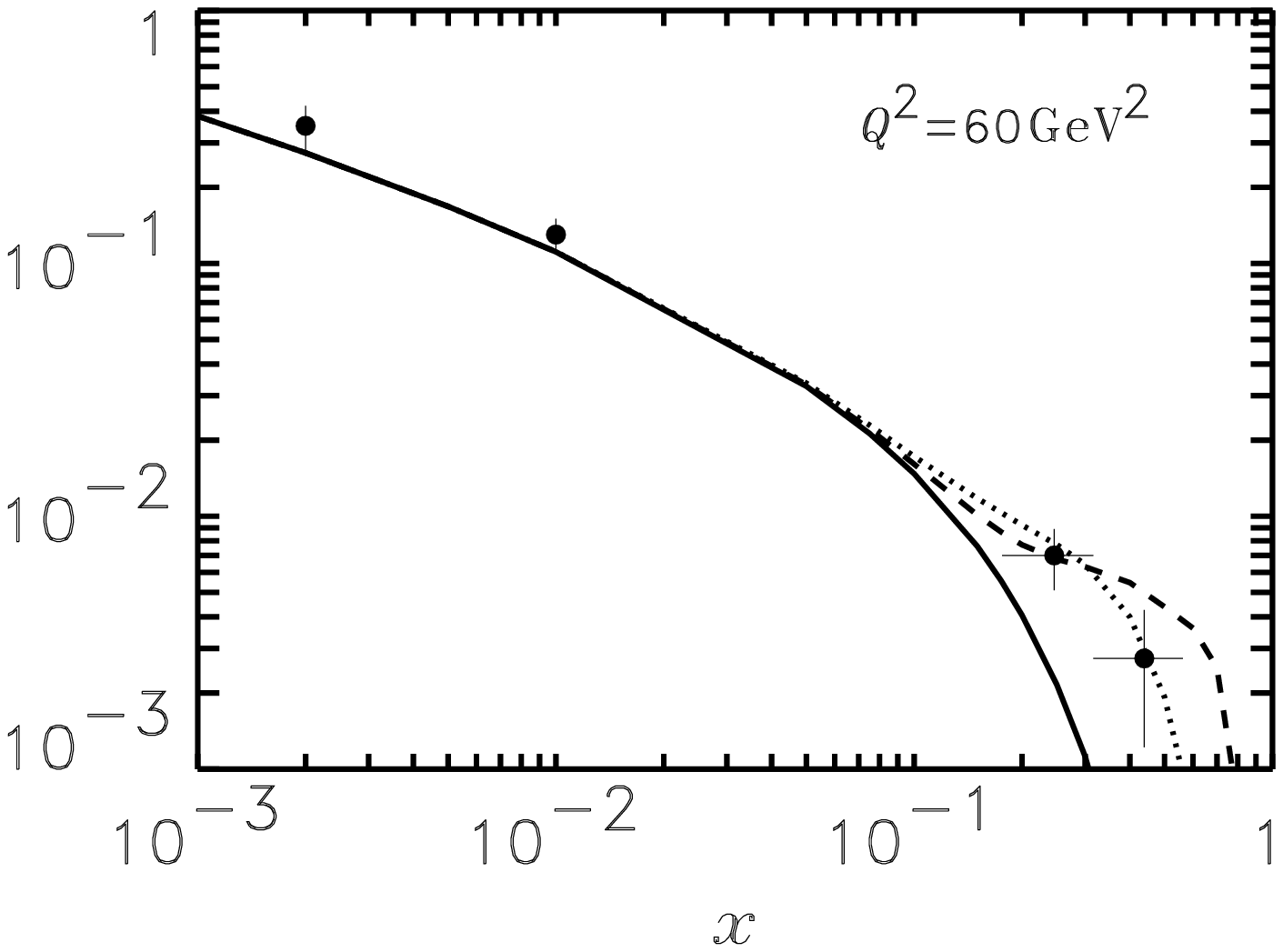,height=7cm}}
\end{picture}}
\vspace*{10cm}
\caption{As in Figure~\protect\ref{C5}, but with $F_2^c$
	calculated according to the FFNS in Eq.(\protect\ref{e5}).}
\label{C7}
\end{figure}

\begin{figure}
\centering{
\begin{picture}(120,150)(180,0)
\epsfig{figure=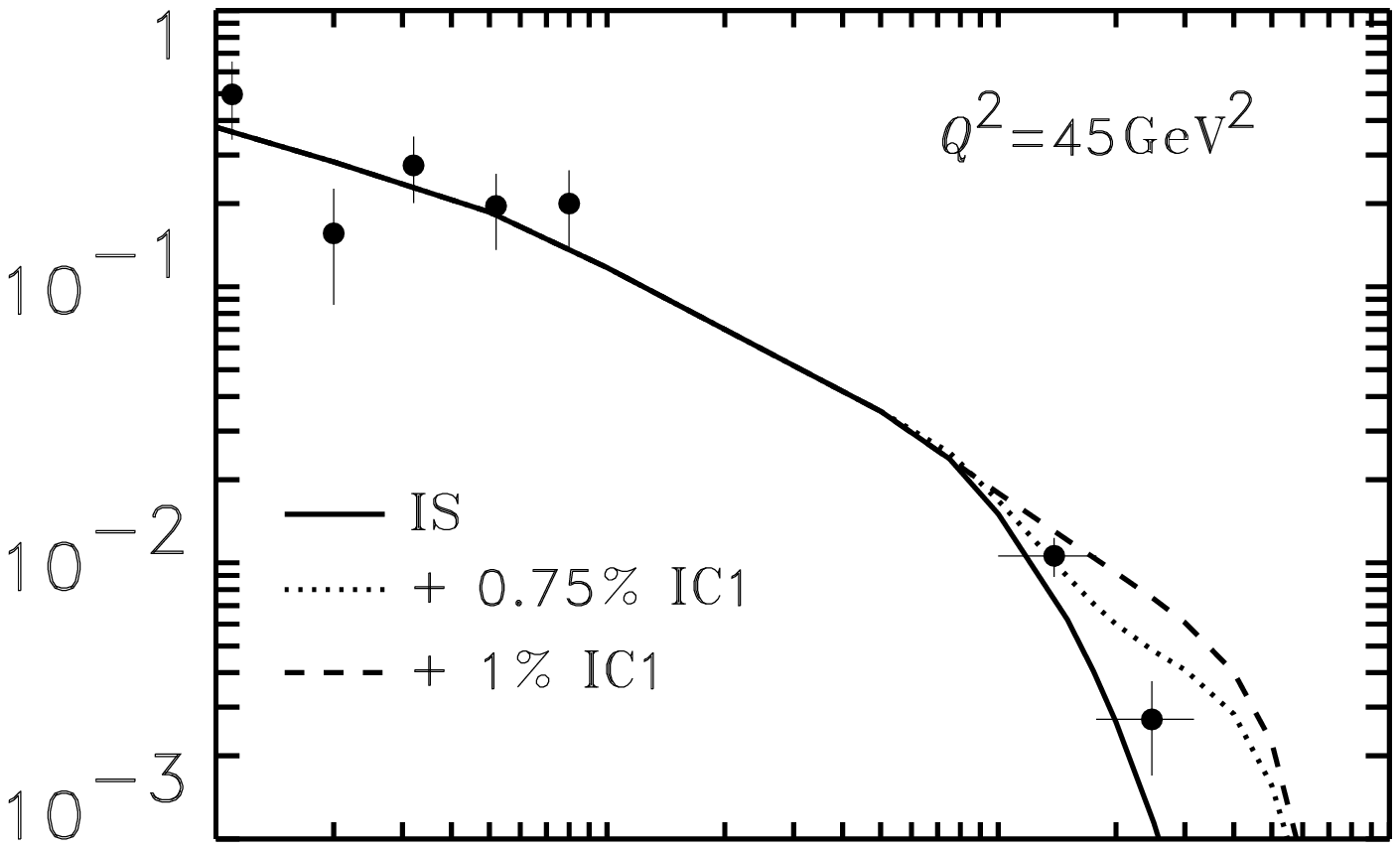,height=7cm}
\put(-279,-135){\epsfig{figure=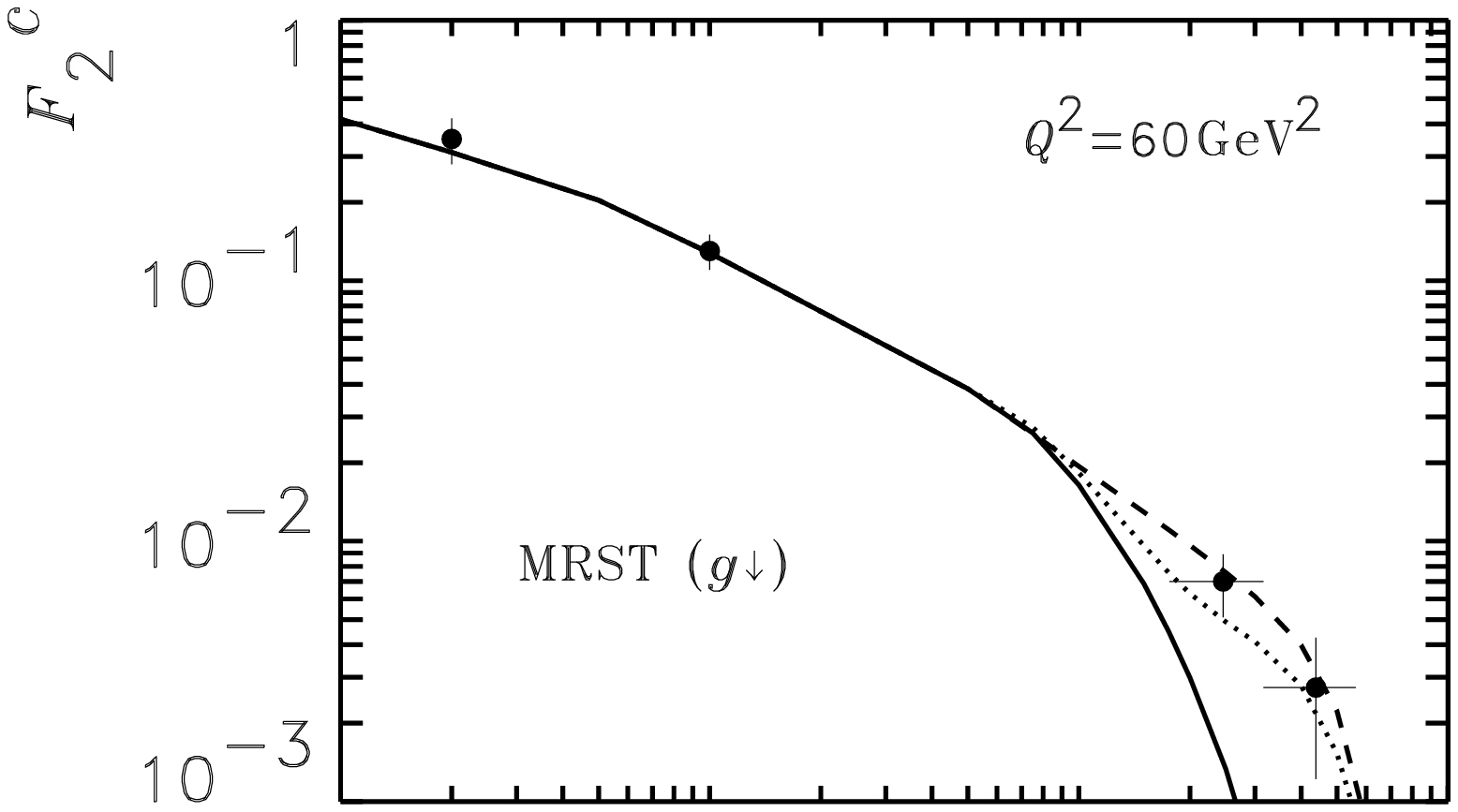,height=7cm}}
\put(-279,-290){\epsfig{figure=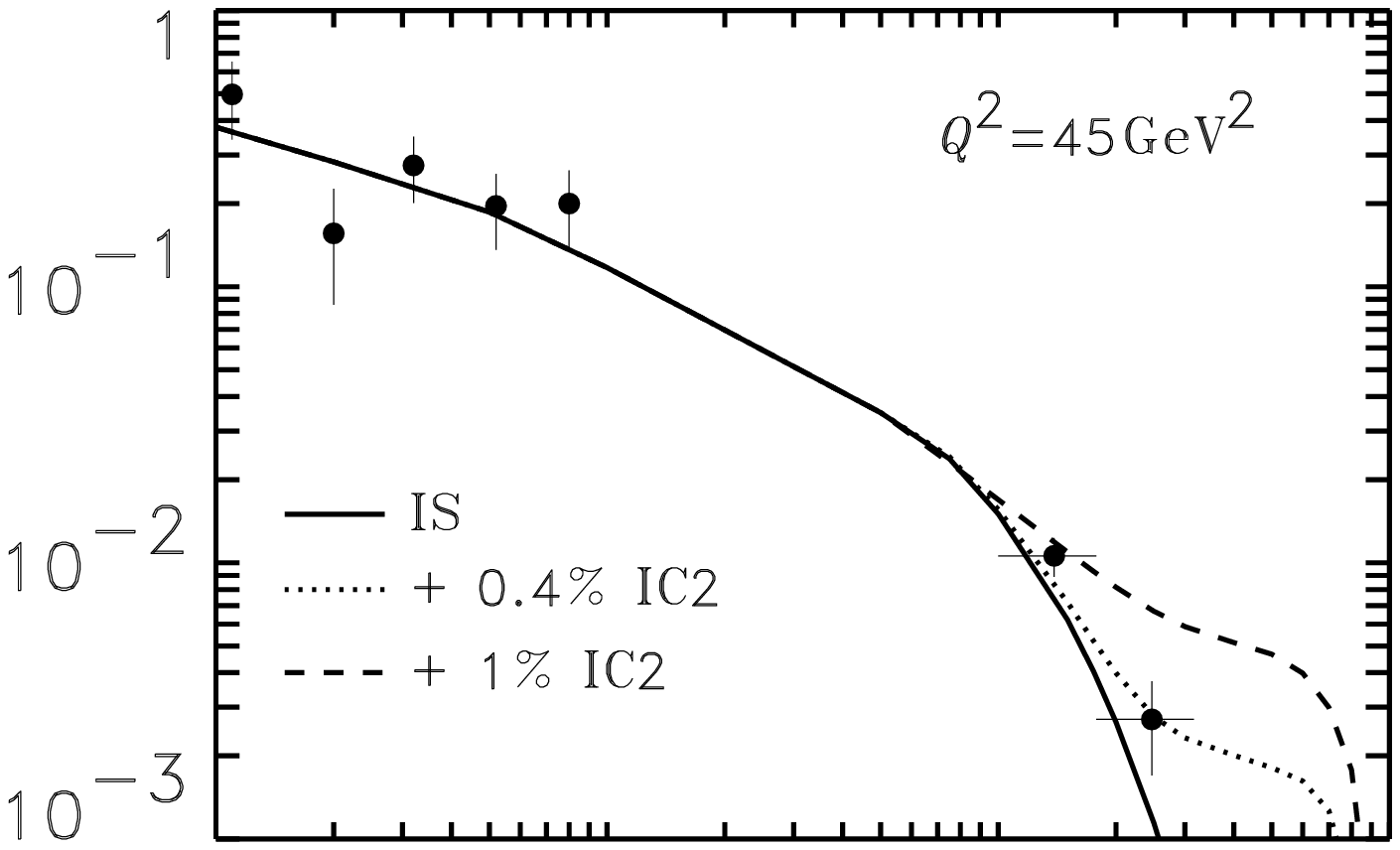,height=7cm}}
\put(-279,-425){\epsfig{figure=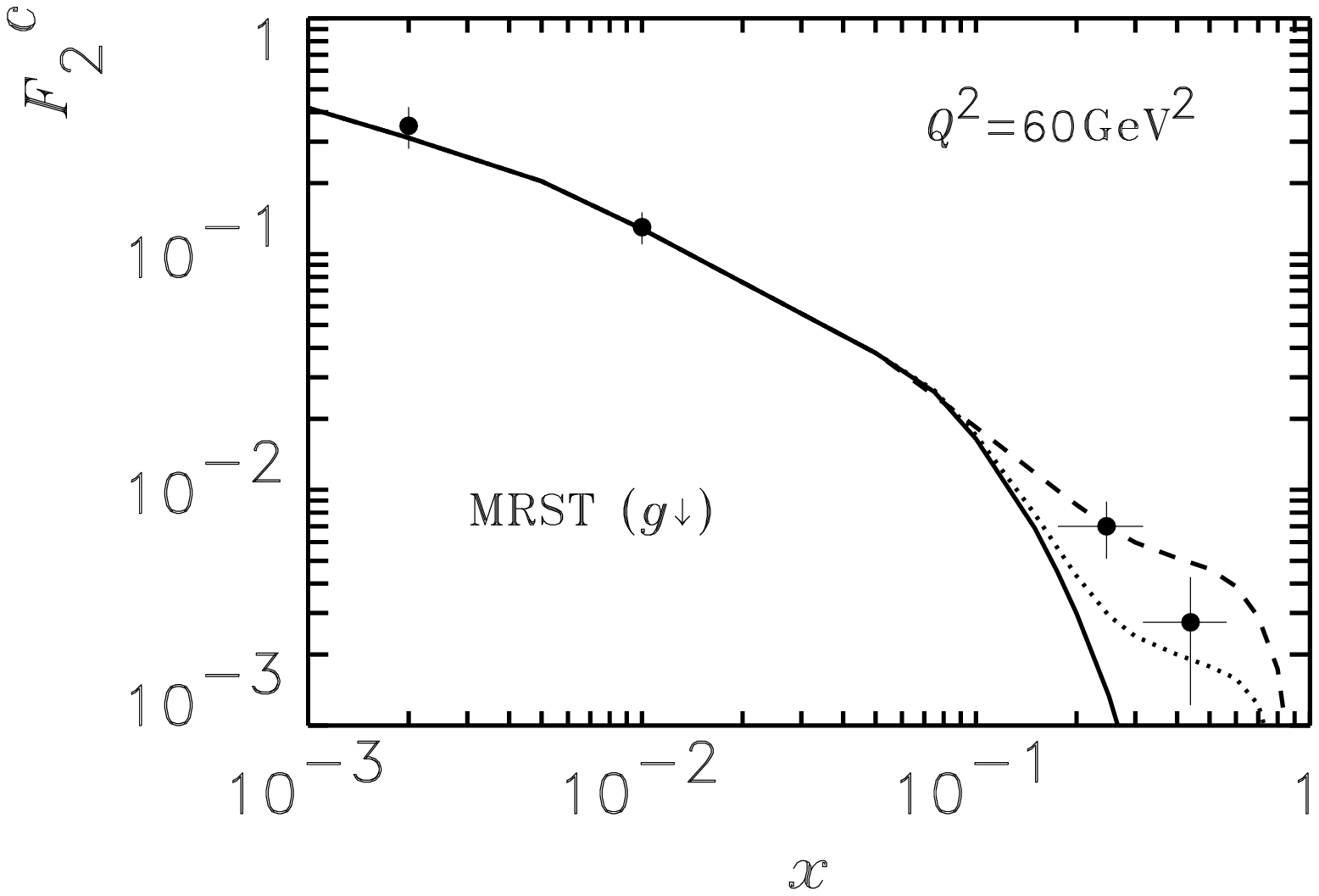,height=7cm}}
\end{picture}}
\vspace*{16cm}
\caption{As in Figure~\protect\ref{C5}, but with different
	normalizations for the IC1 and IC2 model distributions,
	and with the MRST parameterization \protect\cite{MRST}
	with the minimum gluon.}
\label{C8}
\end{figure}

\begin{figure}
\centering{
\begin{picture}(120,150)(180,0)
\epsfig{figure=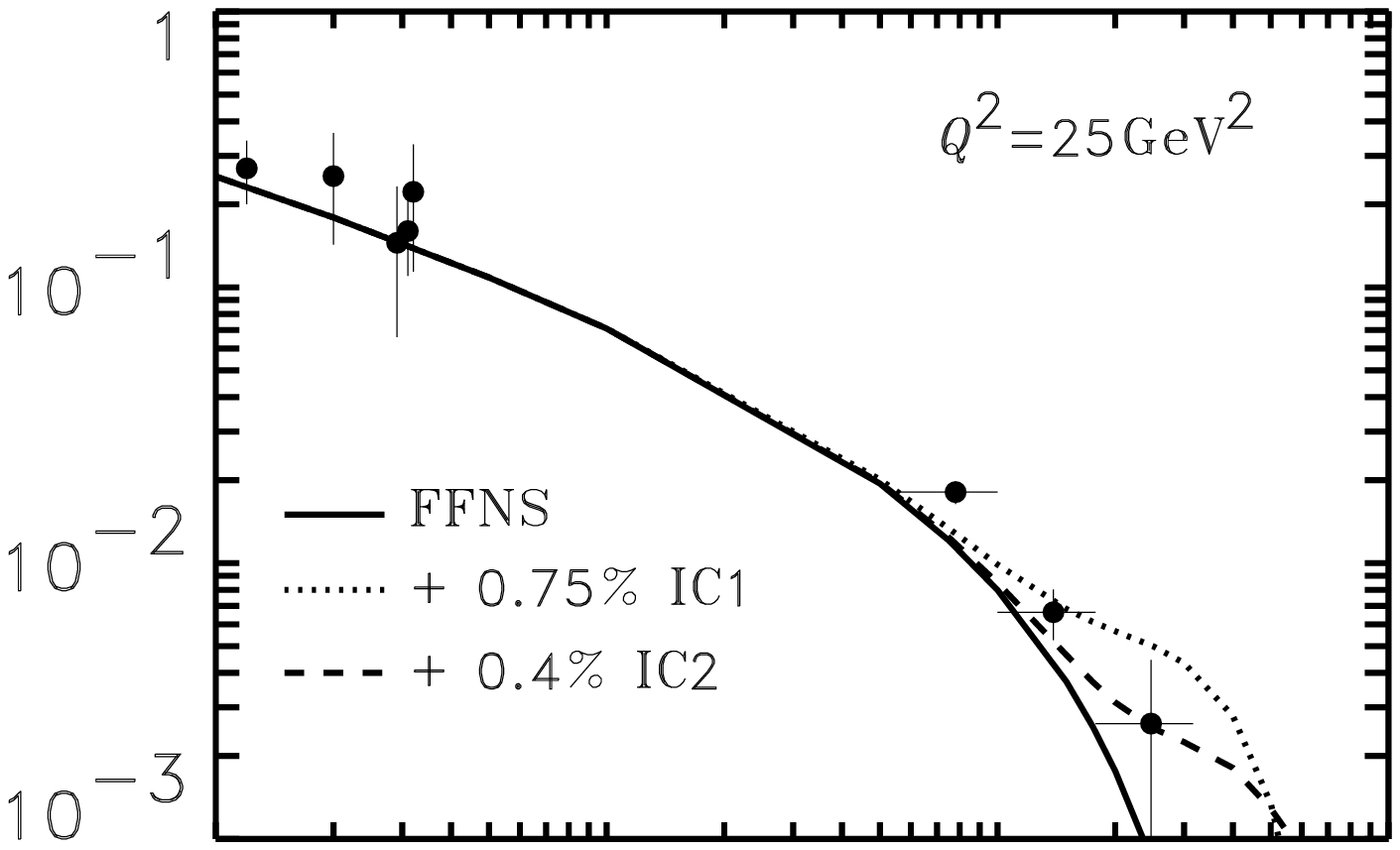,height=7cm}
\put(-279,-135){\epsfig{figure=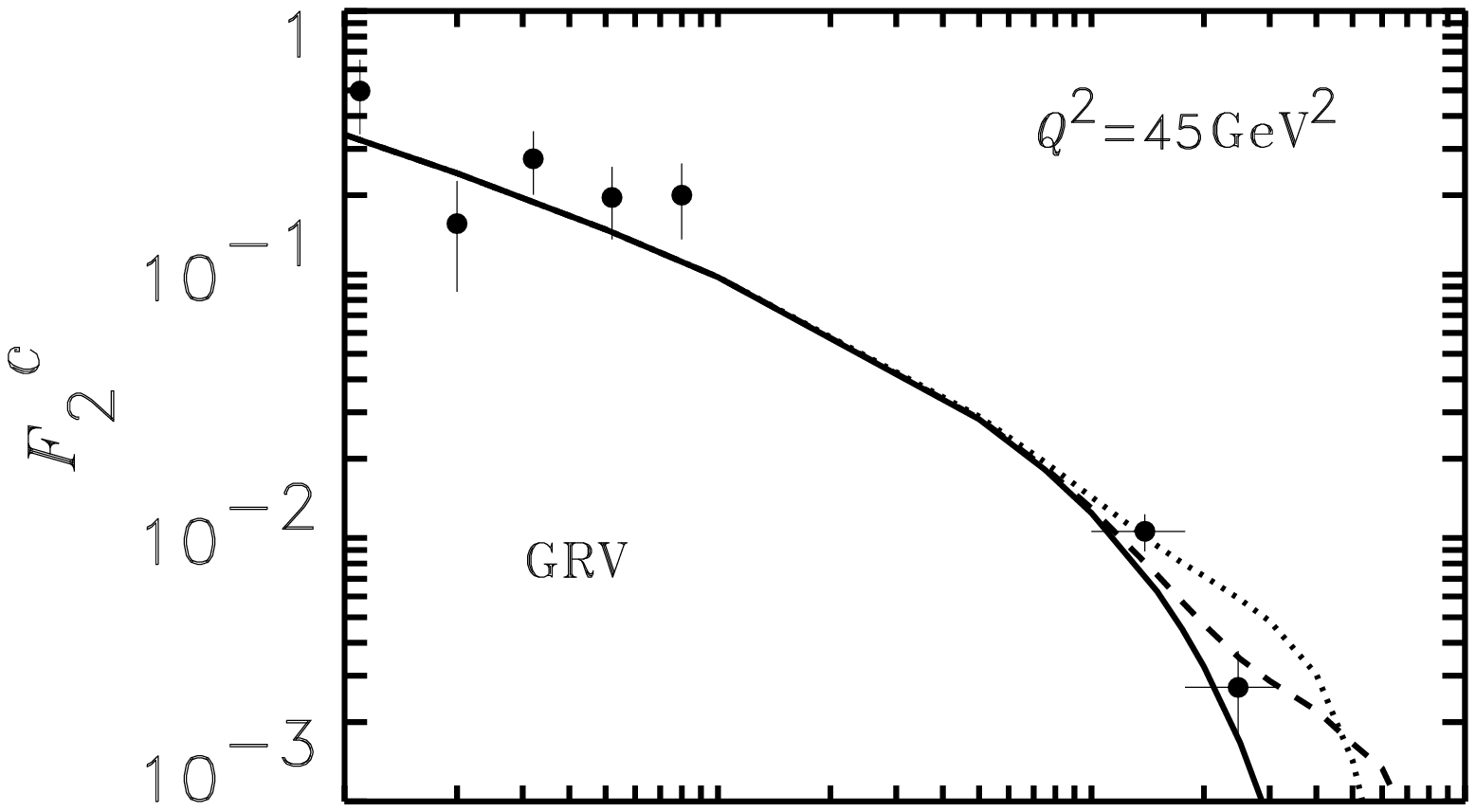,height=7cm}}
\put(-279,-270){\epsfig{figure=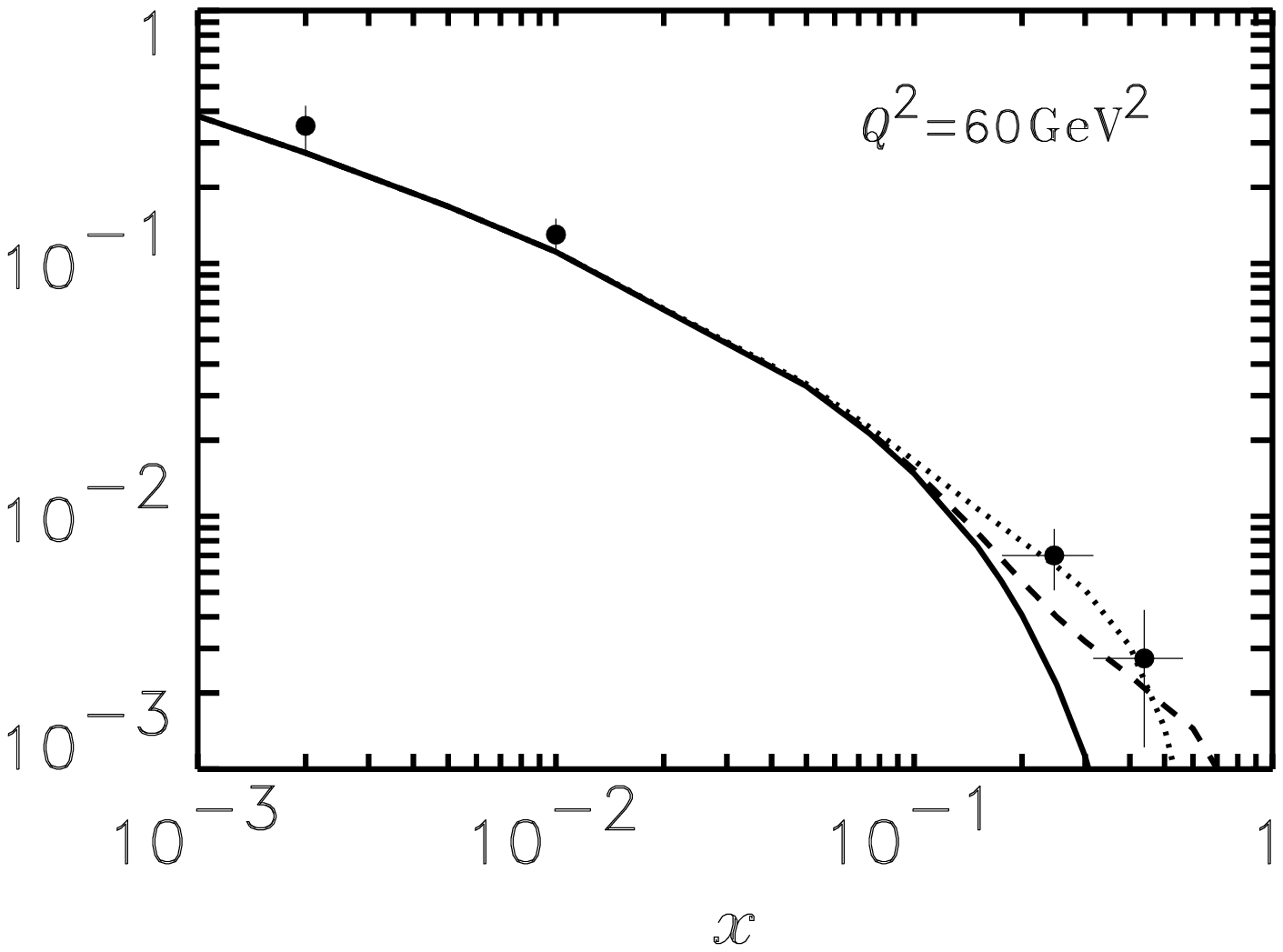,height=7cm}}
\end{picture}}
\vspace*{10cm}
\caption{As in Figure~\protect\ref{C7}, but with different
	normalizations for the IC1 and IC2 components.}
\label{C9}
\end{figure}

\end{document}